\documentclass[runningheads,envcountsame]{llncs}
\usepackage[hidelinks]{hyperref}
\usepackage{amsmath,amsfonts,bm,bbm,ifthen,mathtools}
\usepackage{amssymb}
\usepackage[heavycircles]{stmaryrd}
\usepackage[inline, shortlabels]{enumitem}
\usepackage[sort, noadjust]{cite}
\usepackage{leftidx}
\usepackage{tikz}
\usetikzlibrary{arrows, automata, shapes, calc, petri, arrows.meta}
\tikzset{->, auto, >=stealth', font=\small}
\tikzset{state/.style={shape=circle, draw, fill=white, initial text=, inner sep=.5mm, minimum size=1.5mm}}
\tikzset{accepting/.style=accepting by arrow}
\tikzset{place/.style={shape=circle, draw, minimum size=4mm}}
\tikzset{transition/.style={fill,minimum width=5mm,minimum height=1mm, inner sep=.1mm}}

2

\newcommand*\ie{\textit{i.e.},}
\newcommand*\evord{\dashrightarrow}
\newcommand*\ev{\textup{\textsf{ev}}}
\newcommand*\Nat{\mathbbm{N}}
\newcommand*\arrO[1]{\mathrel{\nearrow^{#1}}}
\newcommand*\arrI[1]{\mathrel{\searrow_{#1}}}
\newcommand*\ess{\textup{\textsf{ess}}}
\newcommand*\loset[1]{\left[\begin{smallmatrix}#1\end{smallmatrix}\right]}
\newcommand*\sloset[1]{\big[\begin{smallmatrix}#1\end{smallmatrix}\big]}
\newcommand*\eg{\textit{e.g.},}
\newcommand*\prepla[1]{\leftidx{^\bullet}{#1}}
\newcommand*\pospla[1]{#1^\bullet}
\newcommand*\prepli[1]{\leftidx{^\circ}{#1}}
\newcommand*\sem[1]{\llbracket #1\rrbracket}
\newcommand*\flatten{\textup{\textsf{flat}}}
\newcommand*\mcal[1]{\mathcal{#1}}
\newcommand*\Inib{\textit{Inib}}
\newcommand*\EL{\textit{EL}}
\newcommand*\starter[2]{\leftidx{_{#2}}{{\hspace*{-.3ex}\uparrow}#1}{}}
\newcommand*\terminator[2]{{#1}{\downarrow}_{#2}}
\newcommand*\ST{\textup{ST}}
\newcommand*\nbtokp[1]{|#1|}

\usepackage[subtle,tracking=normal,wordspacing=tight,mathspacing=tight]{savetrees} 
\makeatletter
\renewcommand\section{\@startsection{section}{1}{\z@}%
  {-12\p@ \@plus -3\p@ \@minus -3\p@}%
  {9\p@ \@plus 3\p@ \@minus 3\p@}%
  {\normalfont\large\bfseries\boldmath
    \rightskip=\z@ \@plus 8em\pretolerance=10000 }}
\makeatother
                    
\begin{document}

\title{Petri Nets and Higher-Dimensional Automata}

\author{%
  Amazigh Amrane\inst1 \and
  Hugo Bazille\inst1 \and
  Uli Fahrenberg\inst1\inst2 \and \\
  Loïc Hélouët\inst2 \and
  Philipp Schlehuber-Caissier\inst3\thanks{%
    Partially funded by the Academic and Research Chair ``Architecture des Systèmes Complexes''
    Dassault Aviation, Naval Group, Dassault Systèmes, KNDS
    France, Agence de l'Innovation de Défense, Institut Polytechnique de Paris}
}

\authorrunning{Amrane, Bazille, Fahrenberg, Hélouët, Schlehuber-Caissier}

\institute{%
  EPITA Research Lab (LRE), France \and
  IRISA \& Inria Rennes, France \and
  SAMOVAR, Télécom SudParis, Institut Polytechnique de Paris, France
}

\maketitle

\begin{abstract}
  Petri nets and their variants are often considered through their interleaved semantics,
  \ie~considering executions where, at each step, a single transition fires.
  This is clearly a miss, as Petri nets are a true concurrency model.
  This paper revisits the semantics of Petri nets as higher-dimensional automata (HDAs) as introduced by van Glabbeek,
  which methodically take concurrency into account.
  We extend the translation to include some common features.
  We consider nets with inhibitor arcs,
  under both concurrent semantics used in the literature,
  and generalized self-modifying nets.
  Finally, we present a tool that implements our translations.
\end{abstract}

\section{Introduction}

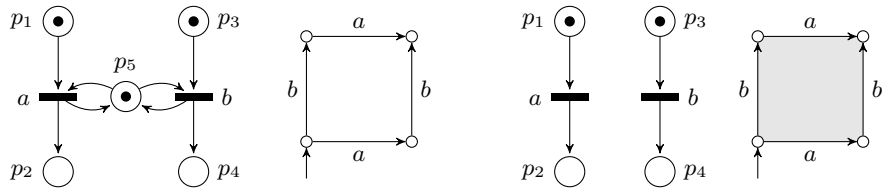
\begin{figure}[tbp]
  \centering
  \begin{tikzpicture}
    \begin{scope}[shift={(-.3,0)}, x=.9cm, y=1cm]
      \node[place, label=above:$p_5$, tokens=1] (5) at (1,-1) {};
      \node[place, label=left:$p_1$, tokens=1] (1) at (0,0) {};
      \node[place, label=left:$p_2$] (2) at (0,-2) {};
      \node[transition, label=left:$\vphantom{b}a$] (t12) at (0,-1) {}
      edge[pre] (1) edge[post] (2)
      edge[pre, bend left] (5) edge[post, bend right] (5);
      \node[place, label=right:$p_3$, tokens=1] (3) at (2,0) {};
      \node[place, label=right:$p_4$] (4) at (2,-2) {};
      \node[transition, label=right:$b$] (t34) at (2,-1) {}
      edge[pre] (3) edge[post] (4)
      edge[pre, bend right] (5) edge[post, bend left] (5);
    \end{scope}
    \begin{scope}[shift={(3,-1.6)}, x=.7cm, y=.7cm]
      \node[state, initial below] (00) at (0,0) {};
      \node[state] (10) at (2,0) {};
      \node[state] (01) at (0,2) {};
      \node[state] (11) at (2,2) {};
      \path (00) edge node[swap] {$a$} (10);
      \path (01) edge node {$a$} (11);
      \path (00) edge node {$b$} (01);
      \path (10) edge node[swap] {$b$} (11);
    \end{scope}
    \begin{scope}[shift={(6.5,0)}, x=.6cm, y=1cm]
      \node[place, label=left:$p_1$, tokens=1] (1) at (0,0) {};
      \node[place, label=left:$p_2$] (2) at (0,-2) {};
      \node[transition, label=left:$\vphantom{b}a$] (t12) at (0,-1) {} edge[pre] (1) edge[post] (2);
      \node[place, label=right:$p_3$, tokens=1] (3) at (2,0) {};
      \node[place, label=right:$p_4$] (4) at (2,-2) {};
      \node[transition, label=right:$b$] (t34) at (2,-1) {} edge[pre] (3) edge[post] (4);
    \end{scope}
    \begin{scope}[shift={(9,-1.6)}, x=.7cm, y=.7cm]
      \path[fill=black!10!white] (0,0) -- (2,0) -- (2,2) -- (0,2) -- (0,0);
      \node[state, initial below] (00) at (0,0) {};
      \node[state] (10) at (2,0) {};
      \node[state] (01) at (0,2) {};
      \node[state] (11) at (2,2) {};
      \path (00) edge node[swap] {$a$} (10);
      \path (01) edge node {$a$} (11);
      \path (00) edge node {$b$} (01);
      \path (10) edge node[swap] {$b$} (11);
    \end{scope}
  \end{tikzpicture}
  \caption{Petri nets and HDAs for interleaving (left) and true concurrency (right).}
  \label{fi:exintro}
\end{figure}

We
revisit the concurrent semantics of Petri nets as higher-dimensional automata (HDAs).
In both Petri nets and HDAs, events may occur simultaneously,
and both formalisms make a distinction between parallel composition $a \parallel b$ and choice $a\cdot b + b\cdot a$.
However, Petri nets are often considered through their interleaving semantics, annihilating this difference.
As an example, Fig.~\ref{fi:exintro} shows two Petri nets and their HDA semantics
(see below for definitions):
on the left, transitions $a$ and $b$ are mutually exclusive and may be executed in any order but not concurrently;
on the right, there is true concurrency between $a$ and $b$, signified by the filled-in square of the HDA semantics.
In interleaving semantics, no distinction is made between the two nets and both give rise to the transition system on the left.

The relations between Petri nets and HDAs were first explored by van~Glab\-beek in~\cite{DBLP:journals/tcs/Glabbeek06},
where an HDA is defined as a labeled precubical set whose
cells are hypercubes of different dimensions.
More recently, \cite{Hdalang}~introduces an event-based setting for HDAs,
defining their cells as totally ordered sets of labeled events.
This framework has led to a number of new developments in the theory of HDAs \cite{%
  DBLP:journals/lmcs/FahrenbergJSZ24,
  DBLP:journals/fuin/FahrenbergZ24,
  DBLP:conf/ictac/AmraneBFZ23,
  DBLP:conf/dlt/AmraneBFF24},
so here we set out to update van Glabbeek's translation to this event-based setting.

Petri nets are a powerful model that can represent infinite systems,
and yet preserve decidability of reachability~\cite{Mayr81} and coverability~\cite{KarpM69}.
Despite their expressiveness, Petri nets miss some features that are essential to represent program executions.
In~\cite{FA73}, the authors introduce inhibitor arcs,
which allow preventing a transition $t$ from firing when a place connected to~$t$ by an inhibitor arc is not empty.
Obviously, this construction allows for the implementation of a zero test,
which makes Petri nets with inhibitor arcs Turing powerful.

We investigate the concurrent semantics of Petri nets with inhibitor arcs,
showing that the a-posteriori semantics of \cite{DBLP:journals/iandc/JanickiK95} gives again rise to HDAs.
For the more liberal a-priori semantics (see again \cite{DBLP:journals/iandc/JanickiK95}) however,
we need to introduce \emph{partial} HDAs in which some cells may be missing,
mimicking the fact that some serialisations of concurrent executions are now forbidden.
We further expand our work to the generalized self-modifying nets of \cite{DBLP:conf/icalp/DufourdFS98},
giving their concurrent semantics as \emph{ST-automata} which themselves generalize partial HDAs.

We have developed a prototype tool which implements the translations from Petri nets to HDAs and from PNIs to partial HDAs.\footnote{%
  See \url{https://gitlabev.imtbs-tsp.eu/philipp.schlehuber-caissier/pn2hda}.}
Our implementation is able to deal with standard, weighted and inhibitor arcs in a modular fashion.

This article is organised as follows.
We begin in Sections~\ref{sec:hda} and~\ref{sec:pn} by recalling HDAs and Petri nets,
focusing on their concurrent semantics which allows several transitions to fire concurrently.
The following sections present our proper contributions.
In Sect.~\ref{sec:pn2hda}, we introduce our translation from Petri nets to HDAs, based on \cite{DBLP:journals/tcs/Glabbeek06}.
To overcome the symmetry of the HDAs thus built,
Sect.~\ref{se:order} introduces an event order which avoids a factorial blow-up in the construction.
We also give several examples to illustrate finer points in HDA semantics.

Then we consider Petri nets with inhibitor arcs in Sect.~\ref{sec:inhibitor},
both under a-posteriori and a-priori semantics,
and generalized self-modifying nets in Sect.~\ref{sec:SMN}.
Section~\ref{sec:implem} presents our implementation.

\section{Higher-Dimensional Automata}
\label{sec:hda}

Higher-dimensional automata (HDAs) extend finite automata
with extra structure which permits to specify independence or concurrency of events.
They consist of \emph{cells} which are connected by \emph{face maps}.
Each cell has a list of events which are active,
and face maps permit to pass from a cell to another in which some events have not yet started or are terminated.

We make this precise.
A \emph{conclist} (\emph{concurrency list}) over an alphabet $\Sigma$
is a tuple $(U, {\evord}, \lambda)$,
consisting of a finite set $U$ (of events),
a strict total order ${\evord}\subseteq U\times U$ (called the {\em event order}),
and a labeling $\lambda: U\to \Sigma$.
Conclists represent labeled events running in parallel.
If no confusion may arise,
we will often refer to conclists by their underlying set only, writing $U$ instead of $(U, {\evord}, \lambda)$,
and do the same for other algebraic structures defined throughout.
Let $\square=\square(\Sigma)$ denote the set of conclists over $\Sigma$.

Conclists $(U_1, {\evord_1}, \lambda_1)$ and $(U_2, {\evord_2}, \lambda_2)$ are \emph{isomorphic}
if there is a mapping $\phi: U_1\to U_2$ such that $a\evord_1 b$ iff $\phi(a)\evord_2 \phi(b)$ and $\lambda_2\circ \phi= \lambda_1$.
Isomorphisms between conclists are unique~\cite{Hdalang},
so we may switch freely between conclists and their isomorphism classes and will do so in the sequel without mention.

\begin{remark}
\label{rem:whyEventOrder}
The event order $\evord$ is important
as a book-keeping device but otherwise carries no computational meaning (see also Sect.~\ref{se:order} below).
It plays a key role in distinguishing between events with the same label (autoconcurrency)
and is needed for uniqueness of conclist isomorphisms.
Conclists without event order are simply multisets,
so conclists are multisets totally ordered by the event order, hence lists or words of $\Sigma^*$;
but we often write them vertically to emphasize that the elements are running in parallel.
Event order goes downwards if not indicated.
\end{remark}

A \emph{precubical set}
\begin{equation*}
  (X, \ev, \{\delta_{A, B; U}\mid U\in \square, A, B\subseteq U, A\cap B=\emptyset\})
\end{equation*}
consists of a set of \emph{cells} $X$
together with a function $\ev: X\to \square$.
For a conclist $U$ we write $X[U]=\{x\in X\mid \ev(x)=U\}$ for the cells of type $U$.
Further, for every $U\in \square$ and $A, B\subseteq U$ with $A\cap B=\emptyset$ there are \emph{face maps}
$\delta_{A, B; U}: X[U]\to X[U\setminus(A\cup B)]$
which satisfy
\begin{equation}
  \label{eq:precid}
  \delta_{C, D; U\setminus(A\cup B)} \delta_{A, B; U}=\delta_{A\cup C, B\cup D; U}
\end{equation}
for every $U\in \square$, $A, B\subseteq U$, and $C, D\subseteq U\setminus(A\cup B)$.

We will omit the extra subscript ``$U$'' in the face maps
and further write $\delta_A^0=\delta_{A, \emptyset}$ and $\delta_B^1=\delta_{\emptyset, B}$.
The \emph{upper} face maps $\delta_B^1$ transform a cell $x$ into one in which the events in $B$ have terminated;
the \emph{lower} face maps $\delta_A^0$ transform $x$ into a cell where the events in $A$ have not yet started.
Every face map $\delta_{A, B}$ can be written as a composition
\begin{equation*}
  \delta_{A, B}=\delta_A^0 \delta_B^1=\delta_B^1 \delta_A^0,
\end{equation*}
and the \emph{precubical identity} \eqref{eq:precid}
expresses  that these transformations commute.

We write $X_n=\{x\in X\mid |\ev(x)|=n\}$ for $n\in \Nat$ and call elements of $X_n$ \emph{$n$-cells}.
The \emph{dimension} of $x\in X$ is $\dim(x)=|\ev(x)|\in \Nat$;
the dimension of $X$ is $\dim(X)=\sup\{\dim(x)\mid x\in X\}\in \Nat\cup\{\infty\}$.
For $k\in \Nat$, the \emph{$k$-truncation} of $X$
is the precubical set $X^{\le k}=\{x\in X\mid \dim(x)\le k\}$
with all cells of dimension higher than $k$ removed.

A \emph{higher-dimensional automaton} (\emph{HDA})
$A=(\Sigma, X, \bot)$ consists of
a finite alphabet $\Sigma$,
a precubical set $X$ on $\Sigma$,
and a subset $\bot\subseteq X$ of initial cells.
(We will not need accepting cells in this work.)
An HDA may be finite or infinite, or even infinite-dimensional.

Computations of HDAs are \emph{paths}, \ie sequences
\begin{equation}
\label{eq_paths}
  \alpha=(x_0, \phi_1, x_1, \dotsc, x_{n-1}, \phi_n, x_n)
\end{equation}
consisting of cells $x_i$ of $X$ and symbols $\phi_i$ which indicate which type of step is taken:
for every $i\in\{1,\dotsc, n\}$, $(x_{i-1}, \phi_i, x_i)$ is either
\begin{itemize}
\item $(\delta^0_A(x_i), \arrO{A}, x_i)$ for $A\subseteq \ev(x_i)$ (an \emph{upstep})
\item or $(x_{i-1}, \arrI{A}, \delta^1_A(x_{i-1}))$ for $A\subseteq \ev(x_{i-1})$ (a \emph{downstep}).
\end{itemize}

Intuitively, a downstep terminates events in a cell, following an upper face map.
This is why downsteps require that $A\subseteq \ev(x_{i-1})$, \ie~events that are terminated belong to the cell. Similarly, 
an upstep starts events by following inverses of lower face maps.
The constraints on upsteps require that $A\subseteq \ev(x_i)$, \ie~the initiated events belong to the next cell after the step.
Both types of steps may be empty.

A cell $x\in X$ is \emph{reachable} if there exists a path $\alpha$ from an initial cell to $x$,
\ie $x_0\in \bot$ and $x_n=x$ in the notation (\ref{eq_paths}) above.
The \emph{essential} part of $X$ is the subset $\ess(X)\subseteq X$ containing only reachable cells.
It is not necessarily an HDA, as some faces may be missing.

\begin{figure}[tbp]
	\centering
	\begin{tikzpicture}[x=.7cm, y=.62cm, every node/.style={transform shape}]
		\begin{scope}[y=.7cm, scale=.9]
			\node[circle,draw=black,fill=blue!20,inner sep=0pt,minimum size=15pt]
			(aa) at (0,0) {$\vphantom{hy}v_1$};
			\node[circle,draw=black,fill=blue!20,inner sep=0pt,minimum size=15pt]
			(ac) at (0,4) {$\vphantom{hy}v_2$};
			\node[circle,draw=black,fill=blue!20,inner sep=0pt,minimum size=15pt]
			(ca) at (4,0) {$\vphantom{hy}v_3$};
			\node[circle,draw=black,fill=blue!20,inner sep=0pt,minimum size=15pt]
			(cc) at (4,4) {$\vphantom{hy}v_4$};
			\node[circle,draw=black,fill=blue!20,inner sep=0pt,minimum size=15pt]
			(ae) at (0,8) {$\vphantom{hy}v_5$};
			\node[circle,draw=black,fill=blue!20,inner sep=0pt,minimum size=15pt]
			(ec) at (8,4) {$\vphantom{hy}v_6$};
			\node[circle,draw=black,fill=blue!20,inner sep=0pt,minimum size=15pt]
			(ce) at (4,8) {$\vphantom{hy}v_7$};
			\node[circle,draw=black,fill=blue!20,inner sep=0pt,minimum size=15pt]
			(ee) at (8,8) {$\vphantom{hy}v_8$};
			\node[circle,draw=black,fill=green!30,inner sep=0pt,minimum size=15pt]
			(ba) at (2,0) {$\vphantom{hy}t_1$};
			\node[circle,draw=black,fill=green!30,inner sep=0pt,minimum size=15pt]
			(bc) at (2,4) {$\vphantom{hy}t_2$};
			\node[circle,draw=black,fill=green!30,inner sep=0pt,minimum size=15pt]
			(ab) at (0,2) {$\vphantom{hy}t_3$};
			\node[circle,draw=black,fill=green!30,inner sep=0pt,minimum size=15pt]
			(ad) at (0,6) {$\vphantom{hy}t_5$};
			\node[circle,draw=black,fill=green!30,inner sep=0pt,minimum size=15pt]
			(be) at (2,8) {$\vphantom{hy}t_6$};
			\node[circle,draw=black,fill=green!30,inner sep=0pt,minimum size=15pt]
			(cd) at (4,6) {$\vphantom{hy}t_8$};
			\node[circle,draw=black,fill=green!30,inner sep=0pt,minimum size=15pt]
			(de) at (6,8) {$\vphantom{hy}t_9$};
			\node[circle,draw=black,fill=green!30,inner sep=0pt,minimum size=15pt]
			(dc) at (6,4) {$\vphantom{hy}t_7$};
			\node[circle,draw=black,fill=green!30,inner sep=0pt,minimum size=15pt]
			(ed) at (8,6) {$\vphantom{hy}t_{10}$};
			\node[circle,draw=black,fill=green!30,inner sep=0pt,minimum size=15pt]
			(cb) at (4,2) {$\vphantom{hy}t_{4}$};
			\node[circle,draw=black,fill=black!20,inner sep=0pt,minimum size=15pt]
			(bb) at (2,2) {$\vphantom{hy}q_1$};
			\node[circle,draw=black,fill=black!20,inner sep=0pt,minimum size=15pt]
			(bd) at (2,6) {$\vphantom{hy}q_2$};
			\node[circle,draw=black,fill=black!20,inner sep=0pt,minimum size=15pt]
			(dd) at (6,6) {$\vphantom{hy}q_3$};
			\path (ba) edge node[above] {$\delta^0_a$} (aa);
			\path (ba) edge node[above] {$\delta^1_a$} (ca);
			\path (bb) edge node[above] {$\delta^0_a$} (ab);
			\path (bb) edge node[above] {$\delta^1_a$} (cb);
			\path (bc) edge node[above] {$\delta^0_a$} (ac);
			\path (bc) edge node[above] {$\delta^1_a$} (cc);
			\path (ab) edge node[left] {$\delta^0_c$} (aa);
			\path (ab) edge node[left] {$\delta^1_c$} (ac);
			\path (bb) edge node[left] {$\delta^0_c$} (ba);
			\path (bb) edge node[left] {$\delta^1_c$} (bc);
			\path (cb) edge node[left] {$\delta^0_c$} (ca);
			\path (cb) edge node[left] {$\delta^1_c$} (cc);
			\path (bb) edge node[above left] {$\delta^1_{ac}\!\!$} (cc);
			\path (bb) edge node[above left] {$\delta^0_{ac}\!\!$} (aa);
			\path (ad) edge node[left] {$\delta^0_d$} (ac);
			\path (ad) edge node[left] {$\delta^1_d$} (ae);
			\path (bd) edge node[left] {$\delta^0_d$} (bc);
			\path (bd) edge node[left] {$\delta^1_d$} (be);
			\path (dc) edge node[above] {$\delta^0_a$} (cc);
			\path (dc) edge node[above] {$\delta^1_a$} (ec);
			\path (bd) edge node[above] {$\delta^0_a$} (ad);
			\path (bd) edge node[above] {$\delta^1_a$} (cd);
			\path (be) edge node[above] {$\delta^0_a$} (ae);
			\path (be) edge node[above] {$\delta^1_a$} (ce);
			\path (bd) edge node[above left] {$\delta^1_{ad}\!\!$} (ce);
			\path (bd) edge node[above left] {$\delta^0_{ad}\!\!$} (ac);
			\path (dd) edge node[above] {$\delta^0_a$} (cd);
			\path (dd) edge node[above] {$\delta^1_a$} (ed);
			\path (de) edge node[above] {$\delta^0_a$} (ce);
			\path (de) edge node[above] {$\delta^1_a$} (ee);
			\path (cd) edge node[left] {$\delta^0_d$} (cc);
			\path (cd) edge node[left] {$\delta^1_d$} (ce);
			\path (dd) edge node[left] {$\delta^0_d$} (dc);
			\path (dd) edge node[left] {$\delta^1_d$} (de);
			\path (ed) edge node[left] {$\delta^0_d$} (ec);
			\path (ed) edge node[left] {$\delta^1_d$} (ee);
			\path (dd) edge node[above left] {$\delta^1_{ad}\!\!$} (ee);
			\path (dd) edge node[above left] {$\delta^0_{ad}\!\!$} (cc);
			\node[below left] at (ab) {$\bot\;\;\;\;$};
		\end{scope}
		\begin{scope}[shift={(-4.7,7)}]
			\node[right] at (9,-7.5) {$X[\emptyset]=\{v_1,\dots, v_8\}$, $X[a]=\{t_1,t_2,t_6,t_7,t_9\}$};
			\node[right] at (9,-6.7) {$X[c]=\{t_3,t_4\}$, $X[d]=\{t_5,t_8,t_{10}\}$};
			\node[right] at (9,-5.9) {$X[\loset{a\\c}]=\{q_1\}$};
			\node[right] at (9,-5.1) {$X[\loset{a\\d}]=\{q_2,q_3\}$};
			\node[right] at (9,-4.3) {$\bot_{X}=\{t_3\}$};
		\end{scope}
		\begin{scope}[shift={(9.8,1.5)}, x=1.3cm, y=1.15cm, scale=.9]
			\filldraw[color=black!15] (0,0)--(2,0)--(2,2)--(0,2)--(0,0);
			\filldraw[color=black!15] (0,2)--(0,4)--(4,4)--(4,2)--(0,2);
			\filldraw (0,0) circle (0.05);
			\filldraw (2,0) circle (0.05);
			\filldraw (0,2) circle (0.05);
			\filldraw (2,2) circle (0.05);
			\filldraw (0,4) circle (0.05);
			\filldraw (4,2) circle (0.05);
			\filldraw (4,4) circle (0.05);
			\filldraw (2,4) circle (0.05);
			\path[line width=.5] (0,0) edge node[below, black] {$\vphantom{b}a$} (1.95,0);
			\path[line width=.5] (0,2) edge node[left, black] {$\vphantom{b}d$} (0,3.95);
			\path[line width=.5] (0,2) edge (1.95,2);
			\path[line width=.5] (2,2) edge node[pos=.6, below, black] {$\vphantom{bg}a$} (3.95,2);
			\path[line width=.5] (2,4) edge (3.95,4);
			\path[line width=.5] (0,0) edge node[pos=.6, left, black] {$\vphantom{bg}c$} (0,1.95);
			\path[line width=.5] (2,0) edge (2,1.95);
			\path[line width=.5] (0,4) edge (1.95,4);
			\path[line width=.5] (2,2) edge (2,3.95);
			\path[line width=.5] (4,2) edge (4,3.95);
			\node[left] at (0,0.9) {$\bot$};
			
			\node[blue,centered] at (0,-0.2) {$v_1$};
			\node[centered, green!50!black] at (1,0.15) {$t_1$};
			\node[blue,centered] at (2,-0.2) {$v_3$};
			\node[centered,blue] at (1.8,2.2) {$v_4$};
			\node[centered,blue] at (-0.2,2) {$v_2$};
			
			\node[centered,blue] at (-0.2,4) {$v_5$};
			\node[centered,blue] at (4.2,2) {$v_6$};
			\node[centered,blue] at (2,4.2) {$v_7$};
			\node[centered,blue] at (4.2,4) {$v_8$};
			
			\node[centered, green!50!black] at (0.2,1.1) {$\vphantom{bg}t_3$};
			\node[centered, green!50!black] at (1.8,1.1) {$\vphantom{bg}t_4$};
			\node[centered, green!50!black] at (1,1.75) {$t_2$};
			\node[centered, green!50!black] at (0.15,3) {$t_5$};
			\node[centered, green!50!black] at (3,2.15) {$t_7$};
			\node[centered, green!50!black] at (3,3.85) {$t_9$};
			\node[centered, green!50!black] at (1.1,3.8) {$\vphantom{bg}t_6$};
			\node[centered, green!50!black] at (2.2,3.1) {$\vphantom{bg}t_8$};
			\node[centered, green!50!black] at (3.8,3.1) {$\vphantom{bg}t_{10}$};
			\node[centered] at (1,1) {$q_1$};
			\node[centered] at (1,3) {$q_2$};
			\node[centered] at (3,3) {$q_3$};
		\end{scope}
	\end{tikzpicture}
	\caption{A two-dimensional HDA $X$ on $\Sigma=\{a, c, d\}$, see Ex.~\ref{ex:hda}.}
	\label{fi:abcube}
\end{figure}
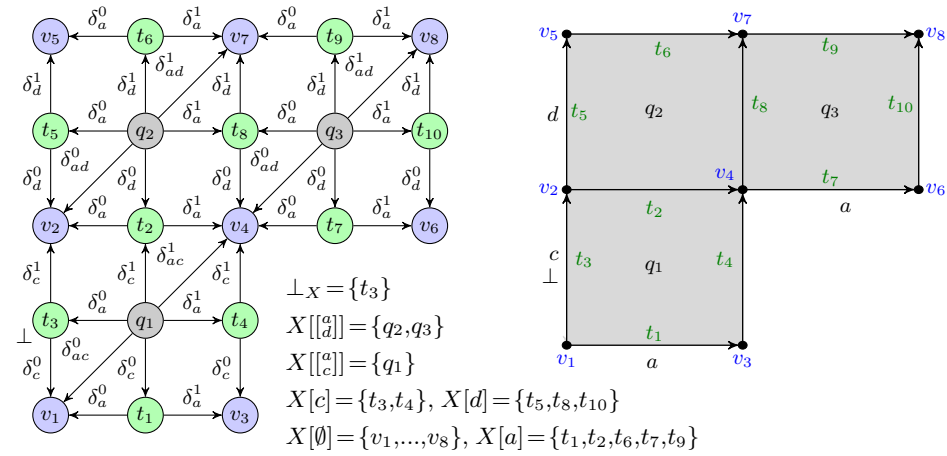

\begin{example}
	\label{ex:hda}
	Figure~\ref{fi:abcube} shows a two-dimensional HDA as a combinatorial object (left)
	and in a geometric realisation (right).
	It consists of
	21 cells:
	states $X_0 = \{v_1,\dots, v_8\}$ in which no event is active ($\ev(v_i) = \emptyset$);
	transitions $X_1 = \{t_1,\dots, t_{10}\}$ in which one event is active (\eg $\ev(t_3) = \ev(t_4) = c$);
	and squares $X_2 = \{q_1, q_2, q_3\}$ with $\ev(q_1) = \loset{a\\c}$ and $\ev(q_2) = \ev(q_3) = \loset{a\\d}$.

	The arrows between cells in the left representation correspond to the face maps connecting them.
	For example, the upper face map $\delta^1_{a c}$ maps $q_1$ to $v_4$
	because the latter is the cell in which the active events $a$ and $c$ of $q_1$ have been terminated.
	On the right, face maps are used to glue cells,
	so that for example $\delta^1_{a c}(q_1)$ is glued to the top right of $q_1$.
	In this and other geometric realisations,
	when we have two concurrent events $a$ and $c$ with $a\evord c$, we will draw $a$ horizontally and $c$ vertically.

	The HDA $X$ of Fig.~\ref{fi:abcube} admits several paths, for example
	$t_3\arrO{a} q_1\arrI{c} t_2 \arrO{d} q_2 \arrI{a} t_8 \arrO{a} q_3 \arrI{ad} v_8$.
	Note that $\ess(X) = X \setminus \{v_1,t_1,v_3\}$ is not an HDA, and $X = X^{\le k}$ for all $k \ge 2$.
\end{example}

\begin{remark}
  We often abuse notation and denote conclists by their labels instead of their events,
  writing for example $\loset{a\\c}$ for the conclist $(\{e_1\evord e_2\}, \lambda:(e_1\mapsto a, e_2\mapsto c))$.
  (We have already done so in the example above.)
  As long as there is no autoconcurrency, this abuse of notation is safe and brings no ambiguity;
  but if we need to assign the same label to several events, we will make events and their labeling explicit in our representation of conclists, writing for instance $\sloset{e_1\mapsto a\\e_2\mapsto c}$ for the above example.
\end{remark}

\section{Petri Nets}
\label{sec:pn}

A \emph{Petri net} $N=(S, T, F)$ consists of
a set of places $S$,
a set of transitions $T$,
with $S\cap T=\emptyset$,
and a weighted flow relation $F: S\times T\cup T\times S\to \Nat$.
A \emph{marking} of $N$ is a function $m: S\to \Nat$.
Such marking is \emph{$k$-bounded} if $m(s)\leq k$ for every place $s$.
It is \emph{bounded} if there is a value $k\in \Nat$ such that $m$ is $k$-bounded.   

Let $X$ be any set. A function $f: X\to \Nat$ is a \emph{multiset}, \ie an extension of sets allowing several instances of each element of $X$.
We introduce some notation for these.
We write $x\in f$ if $f(x)\ge 1$. Given two multisets $f_1,f_2$ over $X$ we will write $f_1\leq f_2$ iff $f_1(x) \leq f_2(x)$ for every element $x\in X$. If $f(x)\in \{0, 1\}$ for all $x$, then $f$ may be seen as a set, and the notation $x\in f$ agrees with the usual one for sets.
The multisets we use will generally be finite in the sense that $\sum_{x\in X} f(x)<\infty$,
and in that case we might use additive notation and write $f=\sum_{x\in X} f(x) x$.
This notation easily applies to markings of Petri nets,
and we will write for instance $m= 2 p_1 + p_4$ for a marking such that $m(p_1)=2, m(p_4)=1$, and $m(p_i) =0$ for any other place $p_i\in S\setminus\{p_1,p_4\}$.

For a transition $t\in T$, the \emph{preset} of $t$ is the multiset $\prepla{t}:S \to \mathbb N$ given by $\prepla{t}(s)=F(s, t)$.
This preset describes how many tokens are consumed in each place when $t$ fires. 
The \emph{postset} of $t$ is the multiset  $\pospla{t}: S\to \Nat$ such that $\pospla{t}(s)=F(t, s)$.
It describes how many tokens are produced in each place of the net when firing $t$.

Petri nets compute by transforming markings.
Their standard semantics is an interleaved semantics,
where states are markings
and a single transition can fire at each step.
Let $m: S\to \Nat$ be a marking and $t\in T$,
then $t$ can \emph{fire} in $m$ if $\prepla{t}\le m$.
Firing $t$ produces a new marking $m'=m-\prepla{t}+\pospla{t}$.

The \emph{reachability graph} (see for example \cite{Desel1998})
of Petri net $N=(S, T, F)$ is the labeled graph $\sem{N}_1=(V, E)$ given by $V=\Nat^S$ and
\begin{equation*}
  E = \{(m, t, m')\in V\times T\times V\mid \prepla{t}\le m, m'=m-\prepla{t}+\pospla{t}\}.
\end{equation*}
(The reason for the subscript $1$ in $\sem{N}_1$ will become clear later.)

In a reachability graph vertices are markings and edges are labeled by the transition which fires.
A computation of a Petri net is a path in its reachability graph.
Note that we use \emph{collective token semantics},
\ie~tokens in $\prepla{t}$ that are consumed by firing $t$ are considered as blind resources.
Petri nets also have an \emph{individual token} semantics~\cite{GR83} where transitions distinguish tokens individually by considering their origin.
This may be used to model realisation of independent processes; but we will not consider it here.

Let $N_1, N_2$ be two Petri nets.
The reachability graphs $\sem{N_1}_1=(V_1, E_1)$ and $\sem{N_2}_1=(V_2, E_2)$ are \emph{isomorphic}, denoted $\sem{N_1}_1 \cong \sem{N_2}_1$,
if there exist bijections $f:V_1\to V_2$ and $g:E_1\to E_2$ such that
for all $e_1=(m_1, t_1, m_1')\in E_1$, $g(e_1)=(m_2,t_2,m_2')$ iff $f(m_1)=m_2$ and $f(m_1')=m_2'$.

Considering Petri nets via their interleaved semantics
misses an important point of the model, namely concurrency. Indeed, it does not allow to distinguish between behaviors where a pair of transitions fire in sequence from behaviors where these transitions are independent and can fire concurrently. 
One way to cope with this issue is to consider executions of Petri nets as {\em processes}~\cite{GR83}, that is, partial orders representing causal dependencies among transitions occurrences. Another possibility is the 
use of a \emph{concurrent step semantics}~\cite{GLT80},
where several transitions are allowed to fire concurrently.
The concurrent step semantics mimics that of the interleaved semantics, but fires multisets of transitions.

For a multiset $U: T\to \Nat$ of transitions
we write $\prepla{U}=\sum_{t\in T} \prepla{t}\, U(t)$
and $\pospla{U}=\sum_{t\in T} \pospla{t}\, U(t)$.
$U$ is \emph{firable} in marking $m$ if $\prepla{U}\le m$. 
The \emph{concurrent step reachability graph} \cite{DBLP:journals/ijfcs/Mukund92}
of Petri net $N=(S, T, F)$ is the labeled graph $\sem{N}_{\textup{CS}}=(V, E)$ given by $V=\Nat^S$ and
\begin{equation}
  \label{eq_conc_def}
  E = \{(m, U, m')\in V\times \Nat^T\times V\mid U\ne \emptyset, \prepla{U}\le m, m'=m-\prepla{U}+\pospla{U}\}.
\end{equation}

\begin{figure}[tbp]
  \centering
  \begin{tikzpicture}[y=.9cm]
    \begin{scope}[shift={(2,0)}, x=.7cm]
      \node[place, label=left:$p_1$, tokens=1] (1) at (0,0) {};
      \node[place, label=left:$p_2$] (2) at (0,-2) {};
      \node[transition, label=left:$\vphantom{b}a$] (t12) at (0,-1) {} edge[pre] (1) edge[post] (2);
      \node[place, label=right:$p_3$, tokens=1] (3) at (2,0) {};
      \node[place, label=right:$p_4$] (4) at (2,-2) {};
      \node[transition, label=right:$b$] (t34) at (2,-1) {} edge[pre] (3) edge[post] (4);
    \end{scope}
    \begin{scope}[shift={(5.8,-2)}, x=1cm]
      \node[state, rectangle, initial] (00) at (0,0) {$p_1+p_3$};
      \node[state, rectangle] (10) at (2,0) {$p_2+p_3$};
      \node[state, rectangle] (01) at (0,2) {$p_1+p_4$};
      \node[state, rectangle] (11) at (2,2) {$p_2+p_4$};
      \path (00) edge node[swap] {$a$} (10);
      \path (01) edge node {$a$} (11);
      \path (00) edge node {$b$} (01);
      \path (10) edge node[swap] {$b$} (11);
    \end{scope}
    \begin{scope}[shift={(10.2,-2)}, x=1cm]
      \node[state, rectangle, initial] (00) at (0,0) {$p_1+p_3$};
      \node[state, rectangle] (10) at (2,0) {$p_2+p_3$};
      \node[state, rectangle] (01) at (0,2) {$p_1+p_4$};
      \node[state, rectangle] (11) at (2,2) {$p_2+p_4$};
      \path (00) edge node[swap] {$a$} (10);
      \path (01) edge node {$a$} (11);
      \path (00) edge node {$b$} (01);
      \path (10) edge node[swap] {$b$} (11);
      \path (00) edge node[swap, sloped] {$a+b$} (11);
    \end{scope}
  \end{tikzpicture}
  \caption{%
    A Petri net $N$ (left);
    the reachability graph $\sem{N}_1$ (middle);
    and its concurrent step reachability graph $\sem{N}_{\textup{CS}}$(right).}
  \label{fi:pn+graph}
\end{figure}
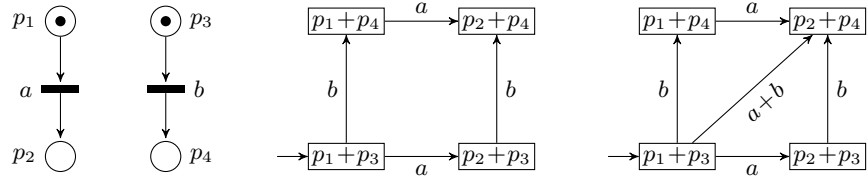

Figure \ref{fi:pn+graph} shows a simple example of a Petri net and its two types of reachability graph.
Note that transitions in $\sem{N}_{\textup{CS}}$ allow multisets of transition rather than only sets,
thus several occurrences of a transition may fire in a concurrent step.
This feature is called autoconcurrency,
and it is well known that allowing autoconcurrency increases the expressive power of Petri nets~\cite{DBLP:conf/concur/Glabbeek05}.
Further, $\sem{N}_{\textup{CS}}$ is \emph{closed under substeps}
in the sense that for all multisets $V \subseteq U$,  if $(m, U, m'')\in E$,
then we also have $(m, V, m')\in E$ and  $(m', U\setminus V, m'')\in E$ for some marking $m'$.

Notice that our definition of Petri nets allows preset-free transitions $t$ with $\prepla{t}=\emptyset$. When a transition $t$ has an empty preset,  then $t$ is firable from any marking. In an interleaved semantics, allowing preset-free transitions does not change expressive power,
so one frequently assumes that $\prepla{t} \neq \emptyset$ for every $t\in T$.   In the setting of a concurrent semantics with autoconcurrency, an arbitrary number of occurrences of each preset-free transitions may fire from any marking, making the transition relation of \eqref{eq_conc_def} of infinite degree.
We will generally allow preset-free transitions in what follows.

\section{From Petri Nets to HDAs}
\label{sec:pn2hda}

We expand the notion of reachability graph to a higher-dimensional automaton (HDA).
The construction is an adaptation of \cite[Def.~9]{DBLP:journals/tcs/Glabbeek06}
to the event-based setting of HDAs introduced in \cite{Hdalang}.

Let $N=(S, T, F)$ be a Petri net.
Let $\square=\square(T)$
and define $X=\Nat^S\times \square$
and $\ev: X\to \square$ by $\ev(m, \tau)=\tau$.
For $x=(m, \tau)\in X[\tau]$ with $\tau=(t_1,\dotsc, t_n)$ non-empty and $i\in\{1,\dotsc, n\}$, define
\begin{align*}
  \delta_{t_i}^0(x) &= (m+\prepla{t_i}, (t_1,\dotsc, t_{i-1}, t_{i+1},\dotsc, t_n)), \\
  \delta_{t_i}^1(x) &= (m+\pospla{t_i}, (t_1,\dotsc, t_{i-1}, t_{i+1},\dotsc, t_n)).
\end{align*}
Using \eqref{eq:precid} to generate the other face maps,
this defines a precubical set $\sem{N}=X$.

The $0$-cells in $X$ are markings of $N$,
and in an $n$-cell of $X$, $n$ transitions of $N$ are running concurrently:
the events of an $n$-cell $(m, \tau)$ are the elements of the conclist $\tau=(t_1,\dotsc, t_n)$ of transitions.
If a transition $t$ appears multiple times in this sequence, then it is autoconcurrent.
Not all precubical sets are in the image of the translation from Petri nets,
see \cite[Fig.~11]{DBLP:journals/tcs/Glabbeek06} for an example.

Note that we translate (unlabeled) Petri nets to HDAs with \emph{labeled} events:
the labels of the events in $\sem{N}$ are the transitions of $N$.
A path in $\sem{N}$ is a computation in $N$ in which concurrent transitions can fire concurrently.

\begin{figure}[tbp]
  \centering
  \begin{tikzpicture}
    \begin{scope}
      \node[align=left, anchor=west] {%
        $X[\emptyset]=\{p_1+p_3, p_2+p_3, p_1+p_4, p_2+p_4\}$ \\
        $X[a]=\{(p_3, a), (p_4, a)\}$ \\
        $X[b]=\{(p_1, b), (p_2, b)\}$ \\
        $X[\loset{a\\b}]=\{(0, \loset{a\\b})\}$ \\
        $X[\loset{b\\a}]=\{(0, \loset{b\\a})\}$
      };
    \end{scope}
    \begin{scope}[shift={(7.5,-1)}, x=1.4cm]
      \path[fill=black!10!white] (0,0) -- (2,0) -- (2,2) -- (0,2) -- (0,0);
      \node[state, rectangle, initial] (00) at (0,0) {$p_1+p_3$};
      \node[state, rectangle] (10) at (2,0) {$p_2+p_3$};
      \node[state, rectangle] (01) at (0,2) {$p_1+p_4$};
      \node[state, rectangle] (11) at (2,2) {$p_2+p_4$};
      \path (00) edge node[swap] {$(p_3, a)$} (10);
      \path (01) edge node {$(p_4, a)$} (11);
      \path (00) edge node {$(p_1, b)$} (01);
      \path (10) edge node[swap] {$(p_2, b)$} (11);
      \node at (1,1) {$(0, \loset{a\\b})$};
    \end{scope}
  \end{tikzpicture}
  \caption{%
    Higher-dimensional automaton (reachable part only) for the Petri net of~Fig.~\ref{fi:pn+graph}.
    Left: sets of cells; right: geometric realisation (not showing $X[\loset{b\\a}]$).}
  \label{fi:pn+hda}
\end{figure}
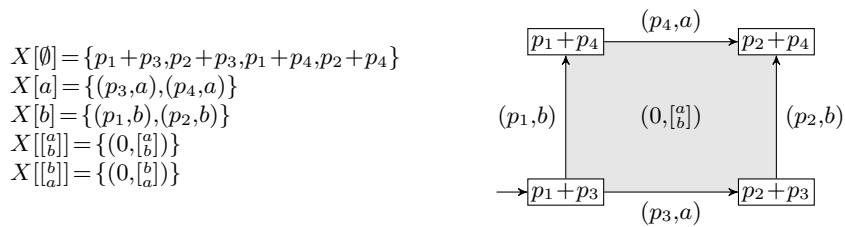

\begin{lemma}
  \label{le:isoReachOneHDA}
  The reachability graph of Petri net $N$ is isomorphic to the $1$-trun\-ca\-tion of $\sem{N}$: $\sem{N}_1\cong \sem{N}^{\le 1}$.
\end{lemma}

In order for this statement to make sense, we must consider $\sem{N}^{\le 1}$ as a graph:
vertices are $0$-cells of $\sem{N}$, and edges triples of the form $(\delta_{t}^0(x), x=(m,t), \delta_{t}^1(x))$.

We can also relate $\sem{N}$ to its concurrent step reachability graph, as follows.
For a sequence $a=(a_1,\dots, a_n)\in \square(\Sigma)$ on some alphabet $\Sigma$
denote by $\textup{pi}(a): \Sigma\to \Nat$ its \emph{Parikh image},
\ie~the multiset given by counting symbols: $\textup{pi}(a)(x)=|\{i\mid a_i=x\}|$.
For a precubical set $X$ on $\Sigma$ define a labeled graph $\flatten(X)=(V, E)$
(the \emph{flattening} of $X$) by $V=X_0$ and $E\subseteq V\times \Nat^\Sigma\times V$ given by
\begin{equation*}
  E = \{(x, U, z)\mid \exists y\in X:
  \delta_{\ev(y)}^0(y)=x, \delta_{\ev(y)}^1(y)=z, \textup{pi}(\ev(y))=U\}.
\end{equation*}
That is, edges in $\flatten(X)$ are labeled by multisets of events for which there exist corresponding cells in $X$,
identifying all permutations in one edge.
We may now reformulate Lem.~\ref{le:isoReachOneHDA} above to $\sem{N}_1\cong \flatten(\sem{N}^{\le 1})$,
and for $\sem{N}_{\textup{CS}}$ the following is clear.

\begin{lemma}
  \label{le:isoCReachHDA}
  The concurrent step reachability graph of a Petri net $N$
  is isomorphic to the flattening of $\sem{N}$:
  $\sem{N}_{\textup{CS}}\cong \flatten(\sem{N})$. \qed
\end{lemma}

Note that under this translation, $\sem{N}_{\textup{CS}}$ being closed under substeps
corresponds to the fact that in $\sem{N}$, all faces of any cell are also present.

A Petri net $N=(S, T, F)$ together with an initial marking $i: S\to \Nat$ is called a \emph{marked Petri net}.
Now $i\in \sem{N}[\emptyset]$,
so this induces an HDA $\sem{N}=(T, X, \bot)$ with $\bot=\{i\}$.

A marked net $N$ is \emph{bounded} if all markings reachable from $i$ in $\sem{N}_1$ are $k-$bounded for some $k\in \mathbb N$.
Obviously, as firable transitions only depend on the current marking, and as the effect of a firing is deterministic, when a marked net is bounded, the reachable part of $\sem{N}_1$, \ie~$\ess(\sem{N}^{\le 1})$ is finite.
However, due to autoconcurrency, this property does not hold for the full $\ess(\sem{N})$, as show in the following example.

\begin{example}
  Let $N=(\emptyset, \{a\}, F)$ be a Petri net with a single transition $a$,
  without places, and with an empty flow relation.
  With empty initial marking, $N$ is bounded.
  Now $\prepla{a}=\emptyset$, so $a$ is firable in arbitrary autoconcurrency.
  In $\ess(\sem{N})$ we get one cell in \emph{every} dimension $n$: the $n$-fold autoconcurrency of $a$.
  That is, $\ess(\sem{N})$ is infinite-dimensional (and hence infinite).
\end{example}

\begin{proposition}
  \label{pr:finiteness}
  If marked Petri net $N$ is bounded and has no preset-free transitions,
  then $\ess(\sem{N})$ is finite.
\end{proposition}

Figure \ref{fi:pn+hda} shows the HDA $\ess(\sem{N})$ for the  Petri net $N$ of Fig.~\ref{fi:pn+graph} with initial marking $i = p_1+p_3$.
In particular, since by construction $0$-cells in $X$ are markings of $N$, $\ess(\sem{N})$ includes all faces, so is an HDA.
(This is a general principle: for any HDA $X$ with $\bot_X\subseteq X_0$, $\ess(X)$ is also an HDA~\cite{DBLP:journals/fuin/FahrenbergZ24}.)

Note that the $2$-dimensional cell $(0, \loset{a\\b})$ corresponds to the edge of $\sem{N}_{\textup{CS}}$ between $p_1+p_3$ and $p_2+p_4$ in Fig.~\ref{fi:pn+graph}.
Actually,
we get \emph{two} $2$-dimensional cells,
one with event $\loset{a\\b}$ and the other with $\loset{b\\a}$.
(We have omitted the second in the geometric realisation.)
This is somewhat unfortunate, as they should denote the same concurrent step $\{a, b\}$.
We will show in Sect.~\ref{se:order} how to fix this problem.

\section{Event Order}
\label{se:order}
\label{sec:example}

The above definition of the HDA $\sem{N}$ is highly symmetric:
for a given marking $m$ there is a cell $(m, \tau)$
for every sequence $\tau=(t_1,\dotsc, t_n)$,
even though \textit{in fine} we are only interested in the \emph{multiset} of concurrently active transitions.
More precisely, for every permutation $\sigma\in \mcal{S}_n$ in the $n^{th}$ symmetric group\footnote{%
  The group of permutations of $\{1,\dots, n\}$.}
there is a cell $(m, \tau\circ \sigma)$.

That is, $X=\sem{N}$ is a \emph{symmetric} precubical set \cite{GrandisM03-Site, DBLP:journals/corr/abs-2409-04612}:
a precubical set equipped with actions $X_n\times \mcal{S}_n\to X_n$ of the symmetric groups
which are consistent with the face maps, see \cite[Sect.~6]{GrandisM03-Site}.

In order to avoid this factorial blow-up,
we may fix an arbitrary (non-strict) total order $\preccurlyeq$ on the transitions in $T$
and then instead of $\square(T)\cong T^*$ consider the set
\begin{equation*}
  T_\preccurlyeq^* = \{(t_1,\dotsc, t_n)\mid \forall i=1,\dotsc, n-1: t_i\preccurlyeq t_{i+1}\}.
\end{equation*}
The definition of the face maps of this reduced $X=\sem{N}$ stays the same,
and $X$ is now a (non-symmetric) precubical set
with one cell for every marking $m$ and every \emph{multiset} of transitions $\tau$.

The order $\preccurlyeq$ on $T$ may be chosen arbitrarily,
and changing it amounts to applying a permutation on $X$ and passing to another, equivalent version of (the symmetrization of) $X$.
Technically speaking, the forgetful functor from symmetric precubical sets to precubical sets
is a \emph{geometric morphism} \cite{Fahrenberg05-thesis, maclane-sheaves} in that it has both a left and a right adjoint;
this is precisely what is needed to be able to say that the order $\preccurlyeq$ on $T$ is arbitrary
and may be chosen and re-chosen at will.

We give some further examples of Petri nets and their HDA semantics,
using a lexicographic order on transitions.

\begin{figure}[tbp]
  \centering
  \begin{tikzpicture}
    \begin{scope}
      \begin{scope}[shift={(5.5,0)}]
        \node[place, label=above:$p_5$, tokens=1] (5) at (1,-1) {};
        \node[place, label=left:$p_1$, tokens=1] (1) at (0,0) {};
        \node[place, label=left:$p_2$] (2) at (0,-2) {};
        \node[transition, label=left:$\vphantom{b}a$] (t12) at (0,-1) {}
        edge[pre] (1) edge[post] (2)
        edge[pre, bend left] (5) edge[post, bend right] (5);
        \node[place, label=right:$p_3$, tokens=1] (3) at (2,0) {};
        \node[place, label=right:$p_4$] (4) at (2,-2) {};
        \node[transition, label=right:$b$] (t34) at (2,-1) {}
        edge[pre] (3) edge[post] (4)
        edge[pre, bend right] (5) edge[post, bend left] (5);
      \end{scope}
    \end{scope}
    \begin{scope}[shift={(0,-3)}]
      \begin{scope}[shift={(1,-1)}]
        \node[align=left, anchor=west] {%
          $X[\emptyset]=\{p_1+p_3+p_5, p_2+p_3+p_5,$ \\
          $\phantom{X[\emptyset]=\{} p_1+p_4+p_5, p_2+p_4+p_5\}$ \\
          $X[a]=\{(p_3, a), (p_4, a)\}$ \\
          $X[b]=\{(p_1, b), (p_2, b)\}$
        };
      \end{scope}
      \begin{scope}[shift={(8,-2)}, x=1.7cm]
        \node[state, rectangle, initial] (00) at (0,0) {$p_1+p_3+p_5$};
        \node[state, rectangle] (10) at (2,0) {$p_2+p_3+p_5$};
        \node[state, rectangle] (01) at (0,2) {$p_1+p_4+p_5$};
        \node[state, rectangle] (11) at (2,2) {$p_2+p_4+p_5$};
        \path (00) edge node[swap] {$(p_3, a)$} (10);
        \path (01) edge node {$(p_4, a)$} (11);
        \path (00) edge node {$(p_1, b)$} (01);
        \path (10) edge node[swap] {$(p_2, b)$} (11);
      \end{scope}
    \end{scope}
  \end{tikzpicture}
  \caption{Petri net (top) and HDA semantics (bottom) of Ex.~\ref{ex:pn2}.}
  \label{fi:pn2}
\end{figure}
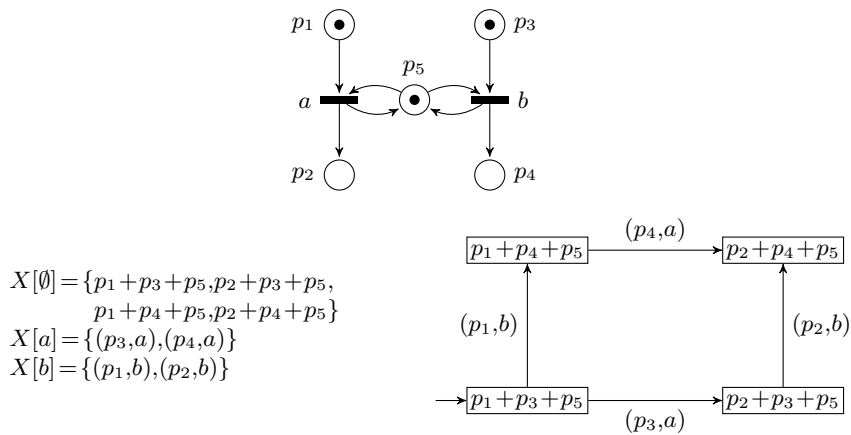

\begin{example}
  \label{ex:pn2}
  Figure \ref{fi:pn2} shows a Petri net $N$ which executes $a$ and $b$ in mutual exclusion.
  We again show the essential part $X=\ess(\sem{N})$;
  the initial cell is $p_1+p_3+p_5$.
  We prove that $X[\loset{a\\b}]=\emptyset$,
  so that $\sem{N}$ is in fact isomorphic to the reachability graph $\sem{N}_1$.
  Assume $x=(m, \loset{a\\b})\in X[\loset{a\\b}]$,
  then $\delta_{a b}^0(x)=m+p_1+p_3+2 p_5$,
  but there is no reachable marking $m'$ with $m'(p_5)=2$.
\end{example}

\begin{figure}[tbp]
  \centering
  \begin{tikzpicture}
    \begin{scope}[shift={(0,0)}, x=.8cm, y=1.2cm]
      \node[place, label=above:$p_1$, tokens=2] (1) at (0,0) {};
      \node[place, label=above:$p_2$, tokens=1] (2) at (1.5,0) {};
      \node[place, label=above:$p_3$, tokens=1] (3) at (3,0) {};
      \node[place, label=left:$p_4$] (4) at (1.5,-2) {};
      \node[transition, label=left:$a$] (t124) at (0.75,-1) {} edge[pre] (1) edge[pre] (2) edge[post] (4);
      \node[transition, label=below:$b$] (t32) at (2.25,-1) {} edge[pre] (3) edge[post] (2);
    \end{scope}
    \begin{scope}[shift={(5,-3)}, x=1.8cm, y=1cm]
      \path[fill=black!10!white] (0,0) -- (2,0) -- (2,4) -- (0,4) -- (0,0);
      \node[state, rectangle, initial] (00) at (0,0) {$2 p_1+p_2+p3$};
      \node[state, rectangle] (10) at (2,0) {$p_1+p_3+p_4$};
      \node[state, rectangle] (01) at (0,2) {$2 p_1+2 p_2$};
      \node[state, rectangle] (11) at (2,2) {$p_1+p_2+p_4$};
      \node[state, rectangle] (02) at (0,4) {$p_1+p_2+p_4$};
      \node[state, rectangle] (12) at (2,4) {$2 p_4$};
      \path (00) edge node[swap] {$(p_1+p_3, a)$} (10);
      \path (01) edge node[swap] {$(p_1+p_2, a)$} (11);
      \path (02) edge node {$(p_4, a)$} (12);
      \path (00) edge node {$(2 p_1+p_2, b)$} (01);
      \path (10) edge node[swap] {$(p_1+p_4, b)$} (11);
      \path (01) edge node {$(p_1+p_2, a)$} (02);
      \path (11) edge node[swap] {$(p_4, a)$} (12);
      \node at (1,1) {$(p_1, \loset{a\\b})$};
      \node at (1,3) {$(0, \loset{a\\a})$};
    \end{scope}
  \end{tikzpicture}
  \caption{%
    Petri net (left) and HDA semantics (right) of Ex.~\ref{ex:contact}.
    Cells with the same label are identified.}
  \label{fi:contact}
\end{figure}
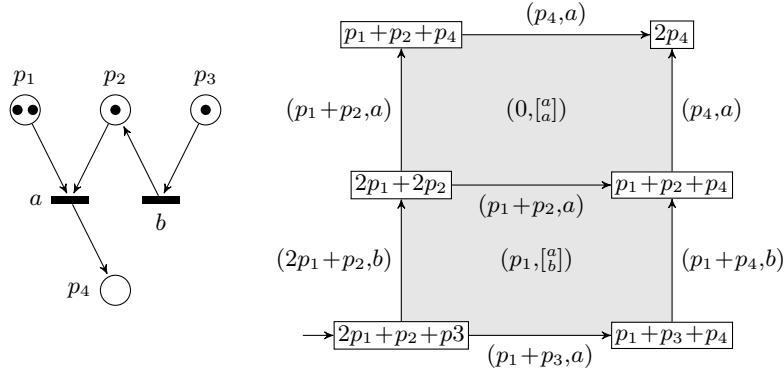

\begin{example}
  \label{ex:contact}
  Figure \ref{fi:contact} shows another Petri net and its HDA semantics.
  Note that there is \emph{contact} between the transitions $a$ and $b$:
  the pre-place $p_2$ is modified when firing $b$.
  Firing $b$ before $a$ enables autoconcurrency of $a$.
  The illustration of the HDA is not quite correct, as the autoconcurrent square $x=(0, \sloset{e_1\mapsto a\\e_2\mapsto a})$
  only has \emph{two} different faces:\footnote{We need to use the extended conclist notation here due to autoconcurrency.}
  using $\cong$ for conclist isomorphism,
  we have $\delta_{e_1}^0(x)=(p_1+p_2, e_2)\cong \delta_{e_2}^0(x)=(p_1+p_2, e_1)$
  and $\delta_{e_1}^1(x)=(p_4, e_2)\cong \delta_{e_2}^1(x)=(p_4, e_1)$.
\end{example}

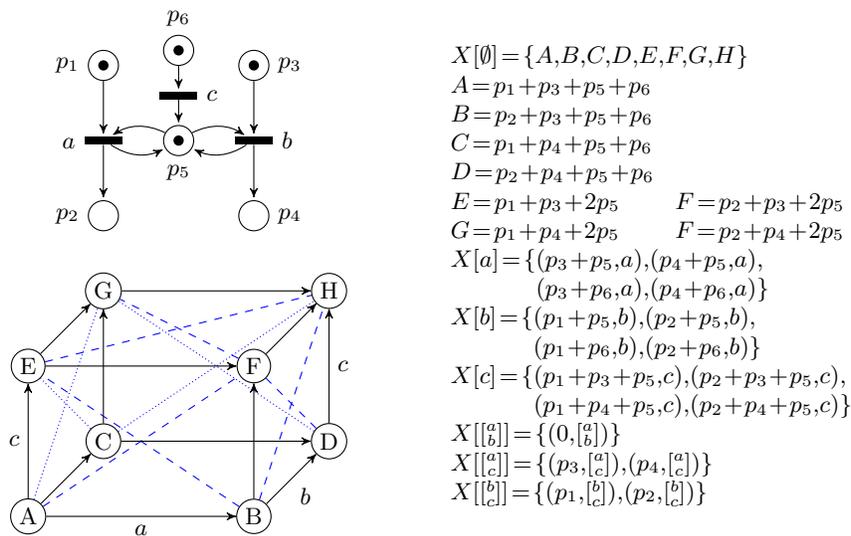
\begin{figure}[tbp]
  \centering
  \begin{tikzpicture}
    \begin{scope}
      \node[place, label=below:$p_5$, tokens=1] (5) at (1,-1) {};
      \node[place, label=left:$p_1$, tokens=1] (1) at (0,0) {};
      \node[place, label=left:$p_2$] (2) at (0,-2) {};
      \node[transition, label=left:$\vphantom{b}a$] (t12) at (0,-1) {}
      edge[pre] (1) edge[post] (2)
      edge[pre, bend left] (5) edge[post, bend right] (5);
      \node[place, label=right:$p_3$, tokens=1] (3) at (2,0) {};
      \node[place, label=right:$p_4$] (4) at (2,-2) {};
      \node[transition, label=right:$b$] (t34) at (2,-1) {}
      edge[pre] (3) edge[post] (4)
      edge[pre, bend right] (5) edge[post, bend left] (5);
      \node[place, label=above:$p_6$, tokens=1] (6) at (1,.2) {};
      \node[transition, label=right:$c$] at (1,-.4) {} edge[pre] (6) edge[post] (5);
    \end{scope}
    \begin{scope}[shift={(4.5,-2.8)}]
      \node[align=left, anchor=west] {%
        $X[\emptyset]=\{A, B, C, D, E, F, G, H\}$ \\
        $A=p_1+p_3+p_5+p_6$ \\ $B=p_2+p_3+p_5+p_6$ \\
        $C=p_1+p_4+p_5+p_6$ \\ $D=p_2+p_4+p_5+p_6$ \\
        $E=p_1+p_3+2 p_5$ \qquad $F=p_2+p_3+2 p_5$ \\
        $G=p_1+p_4+2 p_5$ \qquad $F=p_2+p_4+2 p_5$ \\
        $X[a]=\{(p_3+p_5, a), (p_4+p_5, a),$ \\
        $\phantom{X[a]=\{} (p_3+p_6, a), (p_4+p_6, a)\}$ \\
        $X[b]=\{(p_1+p_5, b), (p_2+p_5, b),$ \\
        $\phantom{X[b]=\{} (p_1+p_6, b), (p_2+p_6, b)\}$ \\
        $X[c]=\{(p_1+p_3+p_5, c), (p_2+p_3+p_5, c),$ \\
        $\phantom{X[c]=\{} (p_1+p_4+p_5, c), (p_2+p_4+p_5, c)\}$ \\
        $X[\loset{a\\b}]=\{(0, \loset{a\\b})\}$ \\
        $X[\loset{a\\c}]=\{(p_3, \loset{a\\c}), (p_4, \loset{a\\c})\}$ \\
        $X[\loset{b\\c}]=\{(p_1, \loset{b\\c}), (p_2, \loset{b\\c})\}$
      };
    \end{scope}
    \begin{scope}[shift={(-1,-6)}]
      \begin{scope}
        \node[state] (x1) at (0,0) {A};
        \node[state] (x2) at (3,0) {B};
        \node[state] (x3) at (1,1) {C};
        \node[state] (x4) at (4,1) {D};
      \end{scope}
      \begin{scope}[shift={(0,2)}]
        \node[state] (y1) at (0,0) {E};
        \node[state] (y2) at (3,0) {F};
        \node[state] (y3) at (1,1) {G};
        \node[state] (y4) at (4,1) {H};
      \end{scope}
      \path[-, blue, dashed] (x1) edge (y2) (x2) edge (y1);
      \path[-, blue, densely dotted] (x1) edge (y3) (x3) edge (y1);
      \path[-, blue, densely dotted] (x3) edge (y4) (x4) edge (y3);
      \path[-, blue, dashed] (x2) edge (y4) (x4) edge (y2);
      \path[-, blue, dashed] (y1) edge (y4) (y2) edge (y3);
      \path (x1) edge node[swap] {$a$} (x2) edge (x3) edge node {$c$} (y1);
      \path (x2) edge node[swap] {$b$} (x4) edge (y2);
      \path (x3) edge (x4) edge (y3);
      \path (x4) edge node[swap] {$c$} (y4);
      \path (y1) edge (y2) edge (y3);
      \path (y2) edge (y4);
      \path (y3) edge (y4);
    \end{scope}
  \end{tikzpicture}
  \caption{%
    Petri net and HDA semantics of Ex.~\ref{ex:pn4}.
    Concurrent squares are indicated using blue crosses instead of filled.}
  \label{fi:pn4}
\end{figure}

\begin{example}
  \label{ex:pn4}
  Figure \ref{fi:pn4} shows a slightly more complicated example,
  where transitions $a$ and $b$ initially are mutually exclusive,
  but then $c$ introduces independence.
  Geometrically this is an empty box without bottom face
  (``Fahrenberg's matchbox'' \cite{DBLP:conf/icalp/DubutGG15}).
  We have $X\!\left[\loset{a\\b\\c}\right]\!=\emptyset$,
  for if $x=\!\left(m, \loset{a\\b\\c}\right)\!\in X\!\left[\loset{a\\b\\c}\right]$, then
  $\delta_{a b c}^0(x)=m+p_1+p_3+2 p_5+p_6$, which is unreachable.
\end{example}

\section{Inhibitor Arcs}
\label{sec:inhibitor}

We now extend our setting to Petri nets with inhibitor arcs.
A \emph{Petri net with inhibitor arcs} (\emph{PNI}) $N=(S, T, F, I)$
consists of a Petri net $(S, T, F)$
and a set $I\subseteq S\times T$ of inhibitor arcs.
We denote by $\prepli{t}=\{s\in S\mid (s, t)\in I\}$
the \emph{inhibitor places} of $t\in T$.

The interleaved semantics for PNIs is as follows.
Tokens in inhibitor places keep transitions from being firable,
so a transition $t\in T$ can fire in marking $m: S\to \Nat$ if $\prepla{t}\le m$ and $\forall s\in \prepli{t}: m(s)=0$.
The reachability graph of a PNI $N=(S, T, F, I)$
is the labeled graph $\sem{N}_1=(V, E)$ given by $V=\Nat^S$ and $E\subseteq V\times T\times V$ with
\begin{equation*}
  E = \{(m, t, m')\mid \prepla{t}\le m, \forall s\in \prepli{t}: m(s)=0, m'=m-\prepla{t}+\pospla{t}\}.
\end{equation*}

PNIs have two different concurrent semantics,
one which disables concurrent steps in which transitions may inhibit each other
and one which does not.
These are called, respectively, \emph{a-posteriori} and \emph{a-priori} semantics in \cite{DBLP:journals/iandc/JanickiK95},
and we treat them both below.
We refer to \cite[Sect.~2]{DBLP:journals/iandc/JanickiK95} for an in-depth discussion of these semantics.

\subsection{Concurrent a-posteriori semantics}

In the a-posteriori semantics, a multiset $U$ of transitions is firable in
$m$ if
\begin{enumerate}[(1)]
\item $\prepla{U}\le m$;
\item for every $t\in U$ and every place $s\in \prepli{t}$, $m(s)=0$;
\item \label{en:busi} for every $t_1, t_2\in U$ such that $t_1=t_2$ implies $U(t_1)\ge 2$,
  $\pospla{t_1}\cap\prepli{t_2}=\emptyset$.
\end{enumerate}

The last condition ensures that $t_1$ cannot produce a token that prevents $t_2$ from firing,
even when $t_1$ and $t_2$ are autoconcurrent in $U$. With this condition,
the transitions in $U$ cannot inhibit \emph{each other}.
One advantage of a-posteriori semantics is that
$\sem{N}_{\textup{CS}}$ is closed under substeps~\cite[Prop.~2.8]{DBLP:journals/tcs/Busi02}.

Similarly to what we did before for Petri nets, we can give an HDA semantics to a PNI $N=(S, T, F, I)$ under a-posteriori concurrent semantics,
by restricting the cells to satisfy conditions (2) and (3) above.
We let again $\square=\square(T)$ and define
\begin{align*}
  X = \{(m, (t_1,\dotsc, t_n))\in \Nat^S\times \square\mid{}
  &\forall i=1,\dotsc, n: \forall s\in \prepli{t_i}: m(s)=0, \\
  &\qquad\forall i\ne j=1,\dotsc, n: \pospla{t_i}\cap \prepli{t_j}=\emptyset\}.
\end{align*}
The rest of the definition of $\sem{N}$ now proceeds as before,
and \cite[Prop.~2.8]{DBLP:journals/tcs/Busi02} ensures that
for any $x\in X$ and any $A\subseteq \ev(x)$, we also have $\delta_A^0(x), \delta_A^1(x)\in X$.

\begin{figure}[tbp]
  \centering
  \begin{tikzpicture}
    \begin{scope}[x=.7cm]
      \node[place, label=left:$p_1$, tokens=1] (1) at (0,0) {};
      \node[place, label=left:$p_2$] (2) at (0,-2) {};
      \node[transition, label=left:$\vphantom{b}a$] (t12) at (0,-1) {} edge[pre] (1) edge[post] (2);
      \node[place, label=right:$p_3$, tokens=1] (3) at (2,0) {};
      \node[place, label=right:$p_4$] (4) at (2,-2) {};
      \node[transition, label=right:$b$] (t34) at (2,-1) {} edge[pre] (3) edge[post] (4);
      \path[-{Circle[open]}] (4) edge (t12);
    \end{scope}
    \begin{scope}[shift={(4,.5)}]
      \node[align=left, anchor=west] {%
        $X[\emptyset]=\{p_1+p_3, p_2+p_3, p_1+p_4, p_2+p_4\}$ \\
        $X[a]=\{(p_3, a)\}$ \\
        $X[b]=\{(p_1, b), (p_2, b)\}$ \\
      };
    \end{scope}
    \begin{scope}[shift={(5,-3)}, x=1.5cm]
      \node[state, rectangle, initial] (00) at (0,0) {$p_1+p_3$};
      \node[state, rectangle] (10) at (2,0) {$p_2+p_3$};
      \node[state, rectangle] (01) at (0,2) {$p_1+p_4$};
      \node[state, rectangle] (11) at (2,2) {$p_2+p_4$};
      \path (00) edge node[swap] {$(p_3, a)$} (10);
      \path (00) edge node {$(p_1, b)$} (01);
      \path (10) edge node[swap] {$(p_2, b)$} (11);
    \end{scope}
  \end{tikzpicture}
  \caption{PNI and HDA semantics of Ex.~\ref{ex:pni-busi}.}
  \label{fi:pni-busi}
\end{figure}
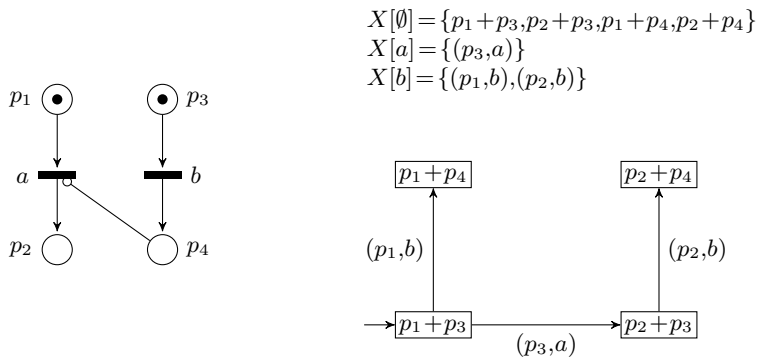

\begin{example}
  \label{ex:pni-busi}
  Figure~\ref{fi:pni-busi} shows the Petri net of Fig.~\ref{fi:pn+graph}
  with an added inhibitor arc from $p_4$ to $a$.
  That is, transition $a$ is disabled when there is a token in $p_4$,
  hence $X[a]$ contains only $(p_3, a)$ but not $(p_4, a)$.
  Compared to Fig.~\ref{fi:pn+hda}, the $2$-cell $(0, \loset{a\\b})$ is also disabled,
  as $\pospla{b}=\prepli{a}=\{p_4\}$:
  there is no concurrency.
\end{example}

\cite{DBLP:journals/tcs/Busi02} also introduces a subclass of PNIs called \emph{primitive systems} defined as follows.
For a marked PNI $N=(S, T, F, I, i)$, let $\Inib(N)=\{ s\in S \mid \exists (s,t)\in I\}$ be the set of places used to inhibit some transition.
Now $N$ is called a \emph{primitive system} if there exists a function $\EL:\Inib(N)\to \Nat$ such that
for all $s\in \Inib(N)$ and all reachable markings $m$ with $m(s)>\EL(s)$,
if $m'$ is reachable from $m$, then
for all $t\in T$ with $\prepla{t}\le m'$,
we have $s \not\in \prepli{t}$.

Intuitively, in primitive systems, when the bound $\EL(s)$ is exceeded in some reachable marking $m$,
then no transition with $s\in \prepli{t}$ is fired in markings that are reachable from $m$.
\cite{DBLP:journals/tcs/Busi02} demonstrates that primitive systems can be simulated by Petri nets (without inhibitor arcs).
However, the author shows that while her construction preserves interleaved semantics, it does not preserve concurrent step semantics;
it is clear that the same is true for our HDA semantics.

\subsection{Concurrent a-priori semantics}

In the more liberal a-priori concurrent semantics, condition \ref{en:busi} of the multiset firing rules is removed.
For intuition, consider again Ex.~\ref{ex:pni-busi} and Fig.~\ref{fi:pni-busi}.
In a-posteriori semantics, the concurrent step $U=\{a,b\}$ is disabled due to condition \ref{en:busi}:
$b$ produces a token in inhibitor place $p_4$ connected to $a$.
This seems rather restrictive: one might argue that \emph{while} the $b$ transition is firing,
it has not yet produced a token in $p_4$,
so it should not prevent from starting the firing of $a$.

On the other hand, if we add the $2$-cell $(0, \loset{a\\b})$ to the semantics,
we are also forced to add its upper face $\delta_b^1((0, \loset{a\\b}))=(p_4, a)$,
given that all faces of cells must be present in HDAs.
Now the cell $(p_4, a)$ would have the upper left vertex $p_1+p_4$ as a lower face,
so semantically that means that we can fire $a$ after firing $b$, which is clearly contrary to what an inhibitor arc should do.
To give proper semantics to PNIs we thus must allow HDAs in which some faces are ``missing''.
These are called \emph{partial HDAs} and have been introduced in~\cite{DBLP:conf/calco/FahrenbergL15, DBLP:conf/fossacs/Dubut19};
we adapt their definition to our event-based setting.

A \emph{partial precubical set}
\begin{equation*}
  (X, \ev, \{\delta_{A, B; U}\mid U\in \square, A, B\subseteq U, A\cap B=\emptyset\})
\end{equation*}
consists of a set of cells $X$
together with a function $\ev: X\to \square$.
Further, for every $U\in \square$ and $A, B\subseteq U$ with $A\cap B=\emptyset$ there are partial face maps
$\delta_{A, B; U}: X[U]\to X[U\setminus(A\cup B)]$
which satisfy
\begin{equation}
  \label{eq:precid-partial}
  \delta_{C, D; U\setminus(A\cup B)} \delta_{A, B; U}\subseteq \delta_{A\cup C, B\cup D; U}
\end{equation}
for every $U\in \square$, $A, B\subseteq U$, and $C, D\subseteq U\setminus(A\cup B)$.
Except for the face maps being partial, this is the same definition as for HDAs in Section~\ref{sec:hda};
we again omit the extra subscripts ``$U$''.
By the notation $\subseteq$ in \eqref{eq:precid-partial} we mean that
if $\delta_{A, B}$ and $\delta_{C, D}$ are defined,
then also $\delta_{A\cup C, B\cup D}$ is defined and equal to the composition $\delta_{C, D} \delta_{A, B}$;
but $\delta_{A\cup C, B\cup D}$ may be defined without one or both of the maps on the left-hand side being defined.

A \emph{partial higher-dimensional automaton}, or \emph{pHDA},
$(\Sigma, X, \bot)$
consists of a partial precubical set $X$ on $\Sigma$ together with a subset $\bot\subseteq X$ of initial cells.

We now give a-priori concurrent semantics to PNIs by translating them to pHDAs.
Let $N=(S, T, F, I)$ be a PNI and $X=\sem{(S, T, F)}$ its standard HDA semantics,
ignoring inhibitor arcs.
Define a subset $X'\subseteq X$ by
\begin{equation*}
  X' = \{(m, (t_1,\dotsc, t_n))\in \Nat^S\times \square\mid \forall i=1,\dotsc, n: \forall s\in \prepli{t_i}: m(s)=0\}
\end{equation*}
and let $\sem{N}=X'$.

Hence a cell $(m, \tau)$ exists in $\sem{N}$ if none of the tokens in $m$ inhibits any transition in $\tau$.
Compared to the contact-free semantics, we leave out Busi's last condition \ref{en:busi},
thus allowing firing of subsets containing pairs of transitions $t_1,t_2$ with $\pospla{t_1}\cap \prepli{t_2}\neq \emptyset$. 

\begin{lemma}
  \label{le:semPartPrec}
  $\sem{N}$ is a partial precubical set.
\end{lemma}

If we extend the definition of flattening and truncation to partial HDAs
and allow the concurrent step semantics to not be closed under substeps,
we have the following analogues of Lemmas \ref{le:isoReachOneHDA} and \ref{le:isoCReachHDA}.

\begin{lemma}
  The reachability graph of PNI $N$ is isomorphic to the $1$-trun\-ca\-tion of $\sem{N}$: $\sem{N}_1\cong \sem{N}^{\le 1}$. \qed
\end{lemma}

\begin{lemma}
  \label{le:isoCReachpHDA}	
  The concurrent step reachability graph of PNI $N$
  is isomorphic to the flattening of $\sem{N}$:
  $\sem{N}_{\textup{CS}}\cong \flatten(\sem{N})$. \qed
\end{lemma}

We also have the following analogue to Prop.~\ref{pr:finiteness}.

\begin{proposition}
  If a marked PNI $N$ is bounded and has no preset-free transitions, then $\ess(\sem{N})$ is finite. \qed
\end{proposition}

\begin{figure}[tbp]
  \centering
  \begin{tikzpicture}[y=.9cm]
    \begin{scope}[x=.65cm]
      \node[place, label=left:$p_1$, tokens=1] (1) at (0,0) {};
      \node[place, label=left:$p_2$] (2) at (0,-2) {};
      \node[transition, label=left:$\vphantom{b}a$] (t12) at (0,-1) {} edge[pre] (1) edge[post] (2);
      \node[place, label=right:$p_3$, tokens=1] (3) at (2,0) {};
      \node[place, label=right:$p_4$] (4) at (2,-2) {};
      \node[transition, label=right:$b$] (t34) at (2,-1) {} edge[pre] (3) edge[post] (4);
      \path[-{Circle[open]}] (4) edge (t12);
    \end{scope}
    \begin{scope}[shift={(5,-2)}, x=1.5cm]
      \path[fill=black!10!white] (0,0) -- (2,0) -- (2,2) -- (0,2) -- (0,0);
      \node[state, rectangle, initial] (00) at (0,0) {$p_1+p_3$};
      \node[state, rectangle] (10) at (2,0) {$p_2+p_3$};
      \node[state, rectangle] (01) at (0,2) {$p_1+p_4$};
      \node[state, rectangle] (11) at (2,2) {$p_2+p_4$};
      \path (00) edge node[swap] {$(p_3, a)$} (10);
      \path (00) edge node {$(p_1, b)$} (01);
      \path (10) edge node[swap] {$(p_2, b)$} (11);
      \node at (1,1) {$(0, \loset{a\\b})$};
    \end{scope}
  \end{tikzpicture}
  \caption{PNI and partial HDA semantics of Ex.~\ref{ex:pni}.}
  \label{fi:pni}
\end{figure}
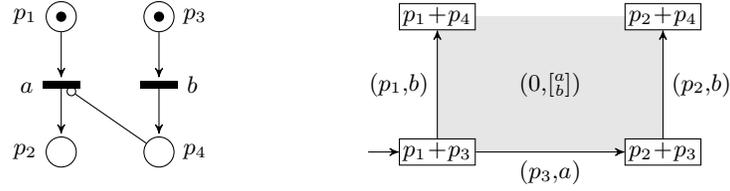

\begin{figure}[tbp]
	\centering
	\begin{tikzpicture}[y=.9cm]
		\begin{scope}[x=.65cm]
			\node[place, label=left:$p_1$, tokens=1] (1) at (0,0) {};
			\node[place, label=left:$p_2$] (2) at (0,-2) {};
			\node[transition, label=left:$\vphantom{b}a$] (t12) at (0,-1) {} edge[pre] (1) edge[post] (2);
			\node[place, label=right:$p_3$, tokens=1] (3) at (2,0) {};
			\node[place, label=right:$p_4$] (4) at (2,-2) {};
			\node[transition, label=right:$b$] (t34) at (2,-1) {} edge[pre] (3) edge[post] (4);
			\path[-{Circle[open]}] (4) edge (t12);
			\path[-{Circle[open]}] (2) edge (t34);
		\end{scope}
		\begin{scope}[shift={(5,-2)}, x=1.5cm]
			\path[fill=black!10!white] (0,0) -- (2,0) -- (2,2) -- (0,2) -- (0,0);
			\node[state, rectangle, initial] (00) at (0,0) {$p_1+p_3$};
			\node[state, rectangle] (10) at (2,0) {$p_2+p_3$};
			\node[state, rectangle] (01) at (0,2) {$p_1+p_4$};
			\node[state, rectangle] (11) at (2,2) {$p_2+p_4$};
			\path (00) edge node[swap] {$(p_3, a)$} (10);
			\path (00) edge node {$(p_1, b)$} (01);
			\node at (1,1) {$(0, \loset{a\\b})$};
		\end{scope}
	\end{tikzpicture}
	\caption{Second PNI and partial HDA semantics of Ex.~\ref{ex:pni}.}
	\label{fi:pnii}
\end{figure}
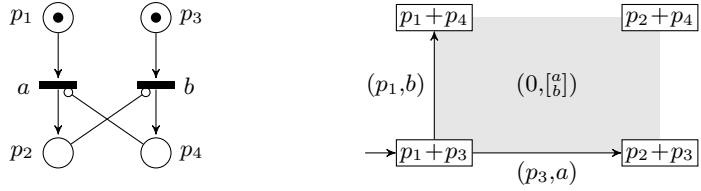

\begin{example}
  \label{ex:pni}
  Consider Fig.~\ref{fi:pni}, where Ex.~\ref{ex:pni-busi} is interpreted with partial HDA semantics, ignoring condition \ref{en:busi}.
  The $2$-cell $(0, \loset{a\\b})$ is now present, but its $\delta_b^1$ face is not.
  There are now
  two paths from $p_1+p_3$ to $p_2+p_4$, passing respectively through $p_2+p_3$ and $(0,\loset{a\\b})$.
  The latter captures reaching marking $p_2+p_4$ from $p_1+p_3$ through the multistep $U = \{a,b\}$ in $\sem{N}_{\textup{CS}}$.

  We may also modify the example by introducing another inhibitor arc from $p_2$ to~$b$,
  see Fig.~\ref{fi:pnii}.
  Then transition $a$ (resp.\ $b$) is disabled when there is a token in $p_4$ (resp.\ $p_2$).
  Again, the corresponding partial HDA contains the $2$-cell $(0,\loset{a\\b})$ even if both its $\delta_a^1$ and $\delta_b^1$ are missing.
  In the a-posteriori semantics, the marking $p_2+p_4$ is now unreachable;
  but in the a-priori semantics,
  there is a path passing through $(0,\loset{a\\b})$ which mimics firing $a$ and $b$ in parallel.
\end{example}

Similarly to \cite[Thm.~5.20]{DBLP:journals/tcs/Busi02}, it can be proven that
the concurrent a-priori semantics of primitive systems may not be simulated by Petri nets without inhibitor arcs.

\section{Self-Modifying Nets}
\label{sec:SMN}

Instead of considering other extensions one by one, we now pass to generalized self-modifying nets
which encompass many other extensions.
Recall \cite{DBLP:conf/icalp/DufourdFS98}
that a \emph{generalized self-modifying net} (\emph{G-net}) $N=(S, T, F)$ consists of
a set of places $S$,
a set of transitions $T$, with $S\cap \Nat=S\cap T=\emptyset$,
and a weighted flow relation $F: S\times T\cup T\times S\to \Nat[S]$.

That is, flow arcs are labeled by \emph{polynomials} in place variables;
contrary to \cite{DBLP:conf/icalp/DufourdFS98} we do not assume that the labels are sums of monomials.
We will propose a concurrent semantics using \emph{ST-graphs}, a generalisation of partial HDAs, see below.

The intuition of the flow polynomials labeling arcs is that when a transition fires, it consumes precisely the number of tokens given by evaluating polynomials of its input arcs, and produces precisely the number of tokens given by evaluating polynomials labeling its output arcs in the current marking.
More precisely,
\begin{itemize}[noitemsep,topsep=0pt]
\item if $F(s, t)=P$ for an arc $(s, t)$, then firing $t$ consumes $P(m)$ tokens from $s$,
  where $m$ is the current marking;
  so the polynomial $P$ is evaluated by replacing its place variables
  with the current number of tokens in the respective places;
\item if $F(t, s)=P$ for an arc $(t, s)$, then firing $t$ produces $P(m)$ tokens in $s$,
  where $m$ is the marking \emph{before} $t$ started firing.
\end{itemize}
That is, in interleaved semantics the new marking when firing a transition $t$ in marking $m$
is given by $m'=m-\prepla{t}(m)+\pospla{t}(m)$:
$\prepla{t}$ is the function $\phi: S\to \Nat[S]$ given by $\phi(s)=F(s,t)$,
so $\prepla{t}(m): S\to \Nat$ is given by $\prepla{t}(m)(s)=\phi(s)(m)$.
Now in concurrent semantics,
other transitions may fire between starting and terminating $t$,
so we will need to remember the marking before firing $t$, see below.

\begin{figure}[tbp]
  \centering
  \begin{tikzpicture}[x=.8cm, y=1.1cm]
    \begin{scope}[shift={(1,0)}]
      \node[place, label=above:$p_1$, tokens=1] (1) at (0,0) {};
      \node[place, label=above:$p_2$, tokens=2] (2) at (1.5,0) {};
      \node[place, label=above:$p_3$, tokens=1] (3) at (3,0) {};
      \node[place, label=left:$p_4$] (4) at (1.5,-2) {};
      \draw[-latex, double] (2) to[bend right=1.5cm] (4);
      \node[transition, label=left:$a$] (t124) at (0.75,-1) {} edge[pre] (1) ;
      \node[transition, label=below:$b$] (t32) at (2.25,-1) {} edge[pre] (3) edge[post] (2);
    \end{scope}
    \begin{scope}[shift={(7,0)}]
      \node[place, label=above:$p_1$, tokens=1] (1) at (0,0) {};
      \node[place, label=above:$p_2$, tokens=2] (2) at (1.5,0) {};
      \node[place, label=above:$p_3$, tokens=1] (3) at (3,0) {};
      \node[place, label=left:$p_4$] (4) at (1.5,-2) {};
      \node[transition, label=left:$a$] (t124) at (0.75,-1) {} edge[pre] (1) edge[pre] node[right] {$p_2$} (2) edge[post] node[right] {$p_2$} (4);
      \node[transition, label=below:$b$] (t32) at (2.25,-1) {} edge[pre] (3) edge[post] (2);
    \end{scope}
  \end{tikzpicture}
  \caption{%
    Petri net with transfer arc (left) and its translation to a G-net (right).
    Annotations ``$p_2$'' at flow arcs
    indicate that they consume and produce the number of tokens present in $p_2$. Otherwise, they consume one token as usual.}
  \label{fi:gnet}
\end{figure}
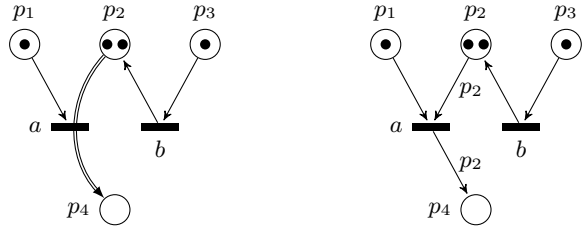

\begin{example}
  \label{ex:pnt}
  G-nets can model \emph{transfer arcs}, a Petri net extension
  which, when firing the associated transition,
  transfers \emph{all} tokens present in the pre-place to the post-place.
  Figure \ref{fi:gnet} shows a simple Petri net $N$ with a transfer arc,
  from $p_2$ through the $a$-transition to $p_4$,
  and its G-net translation.
  (In fact, this is a self-modifying net in the sense of \cite{DBLP:conf/mfcs/Valk78},
  a strict subclass of G-nets.)
  Note that there is contact between the transitions $a$ and $b$,
  and the marking $p_2 + 2 p_4$ reached by firing first $a$ and then $b$
  is not reachable from the marking $p_1 + 3 p_3$ obtained after firing $b$.
  
  The latter is indicated  with a ``broken'' arrow in the intuitive (partial) HDA semantics for $N$:
  \begin{equation*}
    \begin{tikzpicture}[x=1.8cm, y=1cm]
      \path[use as bounding box] (0,-.2) -- (2,-.2) -- (2,2.6) -- (0,2.6) -- (0,-.2);
      \path[fill=black!10!white] (0,0) -- (2,0) -- (2,2) -- (0,2) -- (0,0);
      \node[state, rectangle, initial] (00) at (0,0) {$p_1+2p_2+p_3$};
      \node[state, rectangle] (10) at (2,0) {$p_3+2p_4$};
      \node[state, rectangle] (01) at (0,2) {$p_1+3p_2$};
      \node[state, rectangle] (11) at (2,2) {$p_2+2p_4$};
      \node[state, rectangle] (02) at (2,2.6) {$3p_4$};
      \node[state] (01a) at (.6,2) {};
      \path (00) edge node[swap] {$(p_3, a)$} (10);
      \path (00) edge node {$(p_1+2 p_2, b)$} (01);
      \path (10) edge node[swap] {$(2 p_4, b)$} (11);
      \path (01) edge node {$(0, a)$} (02);
      \path (01a) edge node[swap, pos=.5] {$(p_2, a)$} (11);
      \node at (1,1) {$x=(0, \loset{a\\b})$};
    \end{tikzpicture}
  \end{equation*}
  Transitions $a$ and $b$ can fire concurrently, so we have a square $x\in \sem{N}[\loset{a\\b}]$,
  and from here $b$ can terminate, leading to $\delta_b^1(x)=(p_2, a)$.
  Since $p_2 + 2 p_4$ is not reachable from $p_1+3 p_2$,
  we must have $\delta_a^0((p_2, a))\ne p_1+3 p_2$ if it is defined.
  In fact, this is unreachable, so it is not defined.
  We have a partial HDA, with all cells ``geometrically present'' but one gluing undefined.
\end{example}

In the previous example and Ex.~\ref{ex:pntEx2} below, pHDAs allow to capture the concurrent semantics of G-nets,
however, we conjecture that they are not sufficient in the general case. 
Instead, in order to give concurrent semantics to G-nets,
we introduce a third kind of automaton, closely related to HDAs and partial HDAs but more general.
First, some terminology; again $\Sigma$ denotes an arbitrary alphabet and $\square=\square(\Sigma)$.
A \emph{starter} is a pair $(A, U)$, written $\starter{U}{A}$, consisting of a conclist $U\in \square$ and a subset $A\subseteq U$.
A \emph{terminator} is a pair $(U, B)$, written $\terminator{U}{B}$, consisting of a conclist $U$ and a subset $B\subseteq U$.
The intuition is that these denote actions of starting resp.\ terminating subsets of the events in $U$,
passing from $U\setminus A$ to $U$, resp.\ from $U$ to $U\setminus B$.

Let $\ST=\ST(\Sigma)$ denote the (infinite) set of starters and terminators over~$\Sigma$.
An \emph{ST-graph} is a structure $(\Sigma, Q, E, \lambda)$ consisting of
a set $Q$ of states,
a set $E\subseteq Q\times \ST\times Q$,
and a labeling $\lambda: Q\to \square$
such that for all $(p, x, q)\in E$,
\begin{itemize}
\item if $x=\starter{U}{A}$, then $\lambda(p)=U\setminus A$ and $\lambda(q)=U$;
\item if $x=\terminator{U}{B}$, then $\lambda(p)=U$ and $\lambda(q)=U\setminus B$.
\end{itemize}
ST-\emph{automata}, \ie ST-graphs with initial (and final) states,
have been introduced in \cite{conf/ramics/AmraneBCFZ24} in order to give operational semantics to HDAs.

Now let $N=(S, T, F)$ be a G-net and define an ST-graph $\sem{N}_\ST'=(T, Q', E', \lambda')$ by
$Q'=\Nat^S\times \square\times (\Nat^S)^*$, $\lambda'(m, \tau, \mu)=\tau$,
and
\begin{align*}
  E' = {}
  &\Big\{\Big(\big(m+\prepla{t_i}(\mu_i), (t_1,\dots, t_{i-1}, t_{i+1},\dots, t_n), (\mu_1,\dots, \mu_{i-1}, \mu_{i+1},\dots, \mu_n)\big), \\
  &\qquad \starter{(t_1,\dots, t_n)}{t_i}, \big(m, (t_1,\dots, t_n), \mu\big)\Big)\Big\}
  \\
  {}\cup{}
  &\Big\{\Big(\big(m, (t_1,\dots, t_n), \mu\big), \terminator{(t_1,\dots, t_n)}{t_i}, \\
  &\qquad \big(m+\pospla{t_i}(\mu_i), (t_1,\dots, t_{i-1}, t_{i+1},\dots, t_n), (\mu_1,\dots, \mu_{i-1}, \mu_{i+1},\dots, \mu_n)\big)\Big)\Big\}.
\end{align*}
The intuition is that in a state $x=(m, (t_1,\dotsc, t_n), (\mu_1,\dotsc, \mu_k))\in Q'$,
$n=k$ if $x$ is reachable, and
the marking $\mu_i: S\to \Nat$ is the memory of how the net was marked before transition $t_i$ started firing.

\begin{remark}
  In the construction of $\sem{N}_\ST'$, one may compose successive starting edges to start multiple transitions at the same time,
  similarly for terminating edges.  (See \cite[Sect.~4.4]{conf/ramics/AmraneBCFZ24} for a related construction.)
  Note that concurrency of several transitions is captured
  by starting them one by one in any order before terminating any of them.
\end{remark}

We have primed $\sem{N}_\ST'$ above, as the memory so-defined remembers too much:
in Ex.~\ref{ex:pnt}, transition $a$ only needs to remember the contents of $p_2$ and $b$ should not require memory at all.
We remedy this by introducing a notion of memory equivalence and passing to a quotient.
Say that two pairs $(t, m)$, $(t, m')$ are \emph{memory equivalent},
denoted $(t, m)\sim (t, m')$, if $\prepla{t}(m)=\prepla{t}(m')$ and $\pospla{t}(m)=\pospla{t}(m')$.
Then $m$ and $m'$ have the same \emph{effect} on the net when $t$ is fired.

Now extend $\sim$ to $Q'$ by
$\big(m, (t_1,\dots, t_n), (m_1,\dots, m_n)\big)\sim \big(m, (t_1,\dots, t_n), (m_1',\dots, m_n')\big)$
if $(t_i, m_i)\sim (t_i, m_i')$ for all $i$.
As the memory works by insertion and deletion when starting resp.\ terminating transitions,
$\sim$ is a congruence on the ST-graph $\sem{N}_\ST'$ and we may form the quotient $\sem{N}_\ST=\sem{N}_\ST'/{\sim}$.

Similarly to HDAs, we may define the $1$-truncation of an ST-graph $(\Sigma, Q, E, \lambda)$
as the graph which has as vertices states $q\in Q$ for which $\dim(\lambda(q))=0$
and edges $(q, a, q')$ corresponding to states $x\in Q$ with $\dim(\lambda(x))=1$.
This yields the following analogue of Lem.~\ref{le:isoReachOneHDA};
the relation to the concurrent step reachability graph is
left for future work.

\begin{lemma}
  \label{lem:gnetssta}
  The reachability graph of G-net $N$ is isomorphic to the $1$-trun\-ca\-tion of $\sem{N}_\ST$. \qed
\end{lemma}

We leave open the question of a characterization of G-nets (or subclasses) by means of partial HDAs.
They are sufficient in all our examples.
On the other hand, 
note that the ST-automaton semantics we have given conforms with how the reachable part of the semantics is constructed,
by starting and terminating events one at a time.
(See also Sect.~\ref{sec:implem}.)

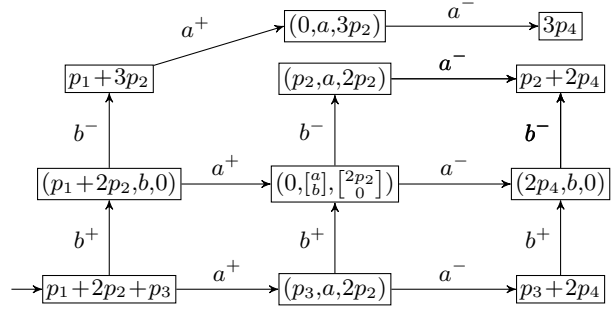
\begin{figure}[tbp]
  \centering
  \begin{tikzpicture}
    \begin{scope}[shift={(0,0)}, x=1.5cm, y=.7cm]
      \node[state, rectangle, initial] (i) at (0,0) {$p_1+2 p_2+p_3$};
      \node[state, rectangle] (a+) at (2,0) {$(p_3, a, 2 p_2)$};
      \node[state, rectangle] (a+a-) at (4,0) {$p_3+2 p_4$};
      \node[state, rectangle] (b+) at (0,2) {$(p_1+2 p_2, b, 0)$};
      \node[state, rectangle] (b+b-) at (0,4) {$p_1+3 p_2$};
      \node[state, rectangle] (a+b+) at (2,2) {$(0, \loset{a\\b}, \sloset{2 p_2\\0})$};
      \node[state, rectangle] (a+a-b+) at (4,2) {$(2 p_4, b, 0)$};
      \node[state, rectangle] (a+b+b-) at (2,4) {$(p_2, a, 2 p_2)$};
      \node[state, rectangle] (b+b-a+) at (2,5) {$(0, a, 3 p_2)$};
      \node[state, rectangle] (a+a-b+b-) at (4,4) {$p_2+2 p_4$};
      \node[state, rectangle] (b+b-a+a-) at (4,5) {$3 p_4$};
      \path (i) edge node {$a^+$} (a+);
      \path (i) edge node {$b^+$} (b+);
      \path (a+) edge node {$a^-$} (a+a-);
      \path (a+) edge node {$b^+$} (a+b+);
      \path (b+) edge node {$b^-$} (b+b-);
      \path (b+) edge node {$a^+$} (a+b+);
      \path (a+a-) edge node {$b^+$} (a+a-b+);
      \path (a+b+) edge node {$a^-$} (a+a-b+);
      \path (a+b+) edge node {$b^-$} (a+b+b-);
      \path (b+b-) edge node {$a^+$} (b+b-a+.west);
      \path (a+a-b+) edge node {$b^-$} (a+a-b+b-);
      \path (a+a-b+) edge node {$b^-$} (a+a-b+b-);
      \path (a+a-b+) edge node {$b^-$} (a+a-b+b-);
      \path (a+b+b-) edge node {$a^-$} (a+a-b+b-);
      \path (a+b+b-) edge node {$a^-$} (a+a-b+b-);
      \path (b+b-a+) edge node {$a^-$} (b+b-a+a-);
    \end{scope}
  \end{tikzpicture}
  \caption{%
    ST-automaton semantics (reachable part only) of the G-net of Fig.~\ref{fi:gnet}, see Ex.~\ref{ex:pnt2sta}.
    For readability, transitions are labeled with starts ($a^+$) and terminations ($a^-$) of actions rather than starters and terminators.}
  \label{fi:pnt2sta}
\end{figure}

\begin{example}
  \label{ex:pnt2sta}
  We continue Ex.~\ref{ex:pnt} by giving $\sem{N}_\ST$ for the transfer net $N$ in Fig.~\ref{fi:pnt2sta},
  indicating memory equivalence classes by representatives.
  We see that, as expected, the sequence $b^+ b^- a^+$ leads to a different state than the other permutations.

  We show some of the calculations; for readability we make a distinction between a variable $p$ appearing in a polynomial
  and the current number of tokens in~$p$, denoting the latter by $\nbtokp{p}$.
  We also write $\prepla t = \sum_s F(s,t) s$ to denote that $t$ fires by consuming $F(s,t)$ tokens from each place $s$ (and similarly for $\pospla t$).

  We have $\prepla{a} = F(p_1,a) p_1 + F(p_2,a) p_2 + F(p_3,a)p_3 = p_1 + \nbtokp{p_2} p_2$ and $\pospla{a} = \nbtokp{p_2} p_4$.
  In addition, the marking before only firing $a$ is $\mu = p_1 + 2p_2 + p_3$.
  Then $\prepla{a}(\mu) = p_1 + \mu(p_2)p_2 = p_1 + 2p_2$.
  Note also that for $\mu' = 2p_2$, for example, $\prepla{a}(\mu) = \prepla{a}(\mu')$ and $\pospla{a}(\mu) = \pospla{a}(\mu')$.
  
  More generally, we have $\prepla{a}(n_1p_1 + n_2p_2 + n_3p_3) = p_1 + n_2 p_2$ for all $n_1,n_2,n_3 \in~\Nat$.
  Thus  $(p_1 + n_2 p_2 + p_3, a^+, (p_3, a, p_1 + n_2 p_2)) \in E'$.
  We also have that $(p_3, a, p_1 + n_2 p_2) \sim (p_3, a, n_2 p_2)$.
  Indeed, since $F(p_1,a)$ and $F(a, p_1)$ are constants, $\prepla{a}(p_1 + n_2 p_2) = \prepla{a}(n_2p_2)$ and $\pospla{a}(p_1 + n_2 p_2) = \pospla{a}(n_2p_2)$.
  
 Thus, for the sequence $a^+ b^+ b^-$, $((p_3, a, 2 p_2), b^+, (0, \loset{a \\ b}, \sloset{2p_2 \\ p_3})) \in E'$ since we had only one token in $p_3$ before firing $b$, but $(0, \loset{a \\ b}, \sloset{2p_2 \\ p_3}) \sim (0, \loset{a \\ b}, \sloset{2p_2 \\ 0})$.
  Finally, $((0, \loset{a \\ b}, \sloset{2p_2 \\ 0}),b^-, (p_2, a, 2p_2)) \in E'$, leading to a different state than $b^+ b^- a^+$.
\end{example}

\begin{figure}[tbp]
  \begin{tikzpicture}
    \begin{scope}[shift={(0,1.7)}, x=.8cm, y=.9cm]
      \node[place, label=above:$p_1$, tokens=1] (1) at (0,0) {};
      \node[place, label=above:$p_2$, tokens=2] (2) at (1.5,0) {};
      \node[place, label=above:$p_3$, tokens=1] (3) at (3,0) {};
      \node[place, label=left:$p_4$] (4) at (.9,-2) {};
      \node[place, label=right:$p_5$] (5) at (2.1,-2) {};
      \draw[-latex, double] (2) to[bend right] (4.north);
      \draw[-latex, double] (2) to[bend left] (5.north);
      \node[transition, label=left:$\vphantom{b}a$] (t124) at (.7,-1) {} edge[pre] (1) ;
      \node[transition, label=right:$b$] (t235) at (2.3,-1) {} edge[pre] (3) ;
    \end{scope}
    \begin{scope}[shift={(4,-3)}, x=1.5cm, y=1cm]
      \node[state, rectangle, initial below] (i) at (0,0) {$p_1+2 p_2+p_3$};
      \node[state, rectangle] (a+) at (2,0) {$(p_3, a, 2 p_2)$};
      \node[state, rectangle] (a+a-) at (4,0) {$p_3+2 p_4$};
      \node[state, rectangle] (b+) at (0,2) {$(p_1, b, 2 p_2)$};
      \node[state, rectangle] (b+b-) at (0,4) {$p_1+2 p_5$};
      \node[state, rectangle] (a+b+) at (2.3,1.7) {$(0, \loset{a\\b}, \sloset{2 p_2\\0})$};
      \node[state, rectangle] (b+a+) at (1.7,2.3) {$(0, \loset{a\\b}, \sloset{0\\2 p_2})$};
      \node[state, rectangle] (a+a-b+) at (4.3,1.7) {$(2 p_4, b, 0)$};
      \node[state, rectangle] (b+a+a-) at (3.7,2.3) {$(0, b, 2 p_2)$};
      \node[state, rectangle] (a+b+b-) at (2.3, 3.7) {$(0, a, 2 p_2)$};
      \node[state, rectangle] (b+a+b-) at (1.7, 4.3) {$(2 p_5, a, 0)$};
      \node[state, rectangle] (a+a-b+b-) at (4.3, 3.7) {$2 p_4$};
      \node[state, rectangle] (b+a+a-b-) at (3.7, 4.3) {$2 p_5$};
      \path (i) edge node {$a^+$} (a+);
      \path (i) edge node {$b^+$} (b+);
      \path (a+) edge node {$a^-$} (a+a-);
      \path (a+.north-|a+b+) edge node {$b^+$} (a+b+);
      \path (b+) edge node {$b^-$} (b+b-);
      \path (b+) edge node {$a^+$} (b+a+);
      \path ($(a+b+.north)!.35!(a+b+.north east)$) edge node[pos=.6] {$b^-$} ($(a+b+b-.south)!.5!(a+b+b-.south east)$);
      \path (b+a+) edge node {$b^-$} (b+a+b-);
      \path (b+b-) edge node {$a^+$} (b+a+b-);
      \path (a+b+b-) edge node[swap, pos=.4] {$a^-$} (a+a-b+b-);
      \path (b+a+b-) edge node {$a^-$} (b+a+a-b-);
      \path (a+a-.north-|a+a-b+) edge node {$b^+$} (a+a-b+);
      \path (a+b+) edge node[swap] {$a^-$} (a+a-b+);
      \path (b+a+) edge node {$a^-$} (b+a+a-);
      \path (a+a-b+) edge node[swap] {$b^-$} (a+a-b+b-);
      \path (b+a+a-) edge node[swap] {$b^-$} (b+a+a-b-);
      \path[densely dashed] (i) edge (a+b+);
      \path[densely dashed] (i) edge (b+a+);
      \node[rotate=37] at (1.2,1.25) {$\{a^+, b^+\}$};
    \end{scope}
    \begin{scope}[shift={(0,-3)}]
      \path[fill=black!10!white] (0,0) -- (2.2,-.4) -- (2.6,1.2) -- (.4,1.6) -- (0,0);
      \path[fill=black!10!white] (0,0) -- (1.8,.4) -- (1.4,2.4) -- (-.4,2) -- (0,0);
      \path[fill=black!20!white] (0,0) -- (1.8,.4) -- (1.6,1.4) -- (.4,1.6) -- (0,0);
      \node[state, rectangle, initial below] (00) at (0,0) {\phantom{$p_1$}\vphantom{$P$}};
      \node[state, rectangle] (10a) at (2.2,-.4) {\phantom{$p_1$}\vphantom{$P$}};
      \node[state, rectangle] (01a) at (.4,1.6) {\phantom{$p_1$}\vphantom{$P$}};
      \node[state, rectangle] (11a) at (2.6,1.2) {$2 p_4$};
      \path (00) edge node[swap, pos=.6] {$a$} (10a);
      \path (01a) edge node[pos=.7] {$a$} (11a);
      \path (10a) edge node[swap] {$b$} (11a);
      \node[state, rectangle] (10b) at (1.8,.4) {\phantom{$p_1$}\vphantom{$P$}};
      \node[state, rectangle] (01b) at (-.4,2) {\phantom{$p_1$}\vphantom{$P$}};
      \node[state, rectangle] (11b) at (1.4,2.4) {$2 p_5$};
      \path (00) edge node {$b$} (01b);
      \path (01b) edge node {$a$} (11b);
      \path (10b) edge node[pos=.4] {$b$} (11b);
    \end{scope}
  \end{tikzpicture}
  \caption{%
    G-net of Ex.~\ref{ex:pntEx2} (top left),
    ST-automaton semantics (right),
    and corresponding partial HDA (bottom left).}
  \label{fi:pntEx2}
\end{figure}
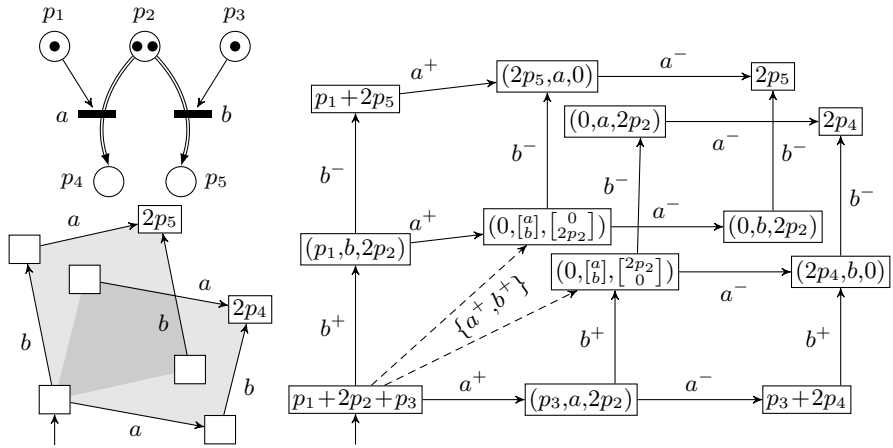

\begin{example}
  \label{ex:pntEx2}
  Figure \ref{fi:pntEx2} shows a G-net containing two transfer arcs which are in conflict
  and the corresponding ST-automaton semantics.
  If we admit that starting $a$ and $b$ at the same time is non-deterministic
  and may lead to any of the two states in the center
  (as indicated by the dashed $\{a^+, b^+\}$-labeled transitions in ST-automaton),
  then this is now a partial HDA: two partial squares labeled $\loset{a\\b}$
  and glued at the initial state.
\end{example}

\section{Implementation}
\label{sec:implem}

We have developed a prototype tool, \texttt{pn2HDA}, written in \texttt{C++} and
implementing our translations from Petri nets to HDAs and from PNIs to partial HDAs.
Our implementation is able to deal with standard, weighted and inhibitor
arcs in a modular fashion and is available at
\url{https://gitlabev.imtbs-tsp.eu/philipp.schlehuber-caissier/pn2hda}.
Our tool is based on previous work by our student Timothée Fragnaud
and on the PNML parser provided by the library Symmetri\footnote{See \url{https://github.com/thorstink/Symmetri}.},
but any other parser could easily replace this task.

As shown above, Petri nets with or without inhibitor arcs (a-posteriori semantics) can be translated into 
the same formalism: partial HDAs.
The implementation reflects this by having a parametrizable (via templates)
representation for Petri nets, which is then used in a generic way to build
the corresponding pHDA.

For this prototype tool, we have chosen an explicit representation of the pHDA
and its face maps as defined at the beginning of section~\ref{sec:pn2hda}. 
That is each reachable cell $x \in X$ is defined as the tuple $(m, \tau)$ 
corresponding to the conclist and the marking.
As event order, the arbitrary total order on transitions introduced in 
section~\ref{se:order}, we have chosen the shortlex order on transition names
(\ie~names are first sorted by their length, and sequences of identical length are sorted according to the lexicographical order).
While this representation is likely not the most efficient, it allows to underpin the 
correctness of our constructions.

Since we cannot display the constructed pHDAs if their dimension is greater
than 3, we have chosen to output them as ST-automata (see Sect.~\ref{sec:SMN}).
For gathering information about the structure of the automaton 
like the number of unique conclists or markings we provide a function $\texttt{get\_csv\_data}$.
In the repository we also provide the PNML files of all Petri nets given
as examples in the paper, as well as a selection of models from the Model Checking Contest\footnote{%
  See \url{https://mcc.lip6.fr}.}.
This is useful for example for gathering statistics about cells of different dimensions,
see Fig.~\ref{fi:implem2}.

\begin{figure}[tbp]
  \centering
  \begin{tikzpicture}
    \begin{scope}[x=.7cm]
      \node[place, label=left:$p_0$, tokens=3] (1) at (0,0) {};
      \node[place, label=left:$p_2$] (2) at (0,-2) {};
      \node[transition, label=left:$\vphantom{b}t0$] (t12) at (0,-1) {} edge[pre] node[right] {2} (1) edge[post] (2);
      \node[place, label=right:$p_1$, tokens=1] (3) at (2,0) {};
      \node[place, label=right:$p_3$] (4) at (2,-2) {};
      \node[transition, label=right:$t1$] (t34) at (2,-1) {} edge[pre] (3) edge[post] (4);
      \path[-{Circle[open]}] (4) edge (t12);
    \end{scope}
    \begin{scope}[shift={(6,-1)}]
      \node {\includegraphics[width=0.6\linewidth]{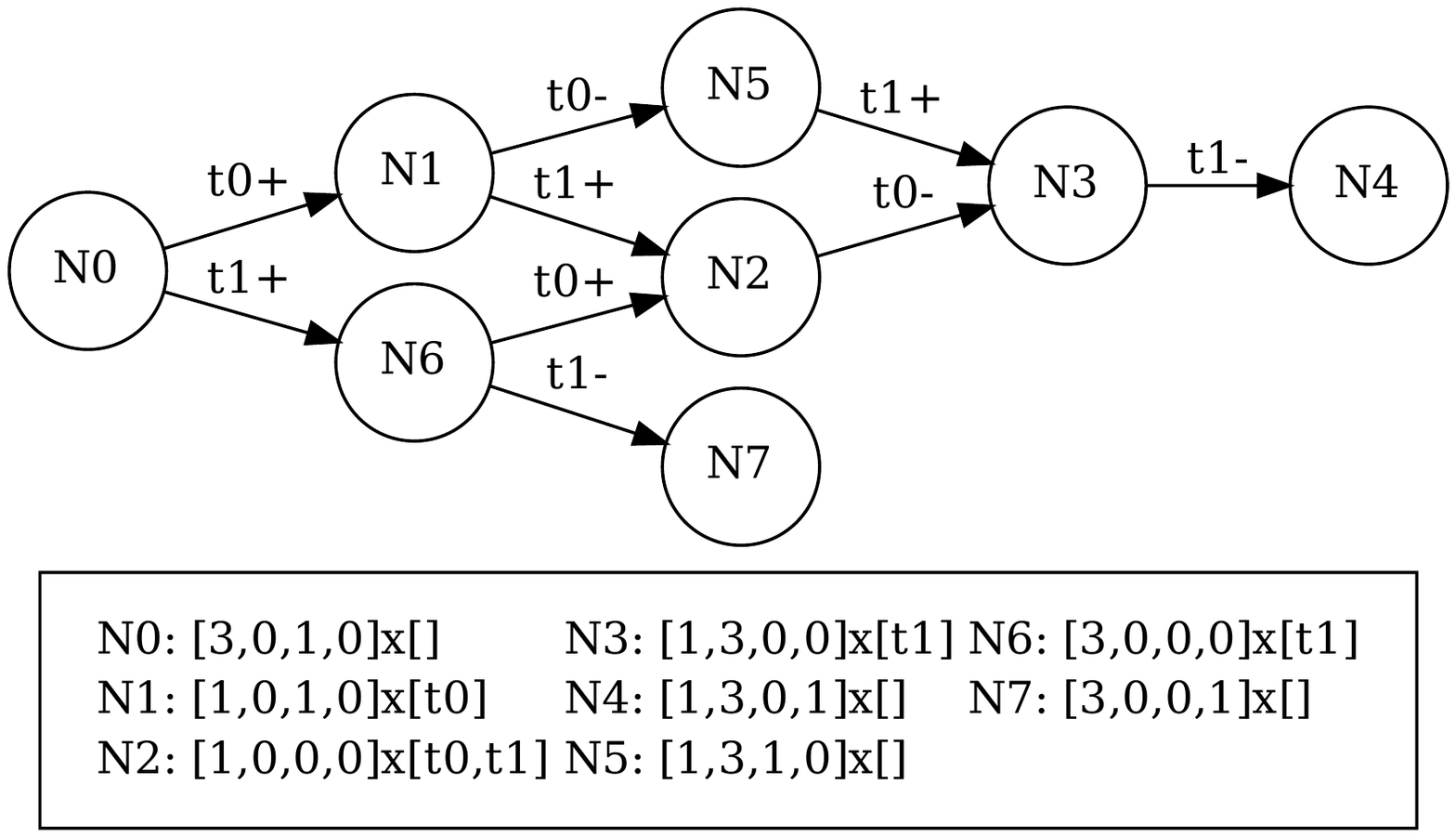}};
    \end{scope}
  \end{tikzpicture}
  \caption{Petri net (left) and pHDA conversion (right): definition of cells (top) and ST-automaton representation (bottom).}
  \label{fi:implem1}
\end{figure}

\begin{figure}[tbp]
  \begin{center}    
  \includegraphics[width=0.75\linewidth]{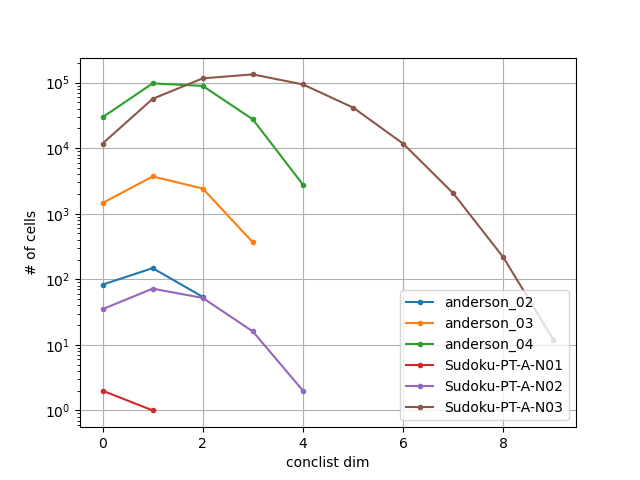}
  \caption{Statistics on cell distribution for some MCC models.}
  \label{fi:implem2}
\end{center}
\end{figure}

\section{Conclusion}
\label{sec:conclusion}

We have seen that Petri nets exhibit a natural concurrent semantics as higher-dimensional automata (HDAs)
which allows methodical reasoning about the finer points of the semantics of Petri nets and their extensions.
The semantics of Petri nets with inhibitors is naturally expressed using partial HDAs
in which some faces may be missing.

We have also given concurrent semantics to generalized self-modifying nets (G-nets)
which encompass many other extensions.
We have given the semantics as ST-automata, a generalization of partial HDAs,
using a notion of memory to store the state of the G-net before the starts of transitions.
Whether or not the semantics may also be given as partial HDAs,
or whether there are interesting subclasses of G-nets which admit partial-HDA semantics, is left open.

Finally, we have presented an implementation of the translations from Petri nets to HDAs
and from Petri nets with inhibitors to partial HDAs.


We believe that pHDA and ST-automaton semantics may provide a unifying framework
for investigating constructions on Petri nets
(such as the removal of inhibitor arcs from primitive systems which we have seen)
and their effects on concurrent semantics.
This should apply to other well-known simplifications such as removing read arcs;
but also for example to unfoldings.

We would also like
to apply our setting to other generalizations of Petri nets.
Introducing concurrent semantics for affine nets \cite{DBLP:conf/fsttcs/BonnetFP12}, for example, appears difficult;
but real-time extensions such as time Petri nets  \cite{merlin1974study} seem natural candidates.
Recent work on higher-dimensional timed automata \cite{%
  conf/apn/AmraneBCF24,
  DBLP:journals/corr/abs-2401-17444}
will be useful in this context.

Finally, we plan to continue to work on our implementation.
We are working on another, implicit, representation of pHDAs
which would avoid creating all reachable cells.
We would also like to extend our tool to time Petri nets
and connect it with Kahl's work on program graphs and homology \cite{DBLP:journals/jact/Kahl24}.\footnote{%
  See \url{https://github.com/twkahl/PG2HDA}.}

\subsection*{Acknowledgements}

We are grateful to Eric Lubat for discussions regarding concurrent semantics,
to Timothée Fragnaud for the initial implementation of the translation from Petri nets to HDAs,
and to Cameron Calk for discovering an error in a earlier version of this work.

\newcommand{\Afirst}[1]{#1} \newcommand{\afirst}[1]{#1}

\appendix
\newpage

\section*{Appendix: proofs}

\begin{proof}[of Lem.~\ref{le:isoReachOneHDA}]
	Let $\sem{N}_1 = (V_N,E_N)$. We can define a bijection $f : V_N \to \sem{N}_0$ between $V_N$ and $0$-cells of $\sem{N}$ as  $f(m) = (m,\emptyset)$ for all $m \in V_N$.
	Then, define $g : E_N \to \sem{N}_1$ such that $g((m_0,t,m_1)) = x = (m,t)$ if $\delta^0(x) = (m_0,\emptyset)$ and $\delta^1(x) = (m_1,\emptyset)$.

	Thus, there exists $e = (m_0,t,m_1) \in E_N$ if and only if $t$ can fire in $m_0$ \ie $\prepla{t} \leq m_0$, and $m_1 = m_0-\prepla{t}+\pospla{t}$ if and only if, by construction of $\sem{N}$, there exists $x \in \sem{N}_1$ and $m= m_0 - \prepla{t} = m_1 - \pospla{t}$ such that $x =  (m,t)$.
    Thus $e = (m_0,t,m_1) \in E_N$ if and only if $g(e) =(f(m_0),(m,t),f(m_1))$.
        \qed
\end{proof}

\begin{proof}[of Prop.~\ref{pr:finiteness}]
	Note first that since $N$ is bounded and $T$ finite there exists a constant $k \in \Nat$ such that for every marking $m$,  $m(p) \leq k$ for all $p \in S$.
	Thus the number of different reachable markings of $\sem{N}_1$ is finite.
	Besides, they correspond to the number of $0$-cells of $\ess(\sem{N})$ by Lem.~\ref{le:isoReachOneHDA}.
	Now, by Lem.~\ref{le:isoCReachHDA}, proving that the number of edges  of the reachable part of $\sem{N}_{\textup{CS}}$ is finite proves the proposition.
	Indeed, each edge $(m,U,m')$ corresponds to a $|U|$-cell of $\sem{N}$, $U \in \square(T)$.
	
	Since each $t \in T$ is constrained, \ie not preset-free, each $t$ fires in some bounded marking $m$ only if $\prepla{t} \leq m$.
	Thus, for each reachable marking $m$ there are finitely many $U \in \square(T)$ such that $\prepla{U} \leq m$, and each such $U$ leads to a unique marking $m' = m - \prepla{U} + \pospla{U}$.
	Note that these $U$ may contain autoconcurrency but are finite since $m$ is bounded.
	In addition, if $(m,U,m')$ is an edge of $\sem{N}_{\textup{CS}}$, then there are $|U|!$ $|U|$-cells: $\{(m - \prepla{U'} = m' - \pospla{U'}, U')\mid U' \text{ a permutation of } U\}$ in $\sem{N}$ for which $(m,\emptyset)$ is a lower face and $(m',\emptyset)$ an upper face  but only one edge $((m,\emptyset),U,(m',\emptyset))$ in $\flatten(\sem{N})$ by construction.
	Finally, since the reachable part of $\sem{N}_{\textup{CS}}$ is finite and contains finitely many edges, $\ess(\sem{N})$ is finite by Lem.~\ref{le:isoCReachHDA}.
        \qed
\end{proof}

\begin{proof}[of Lem.~\ref{le:semPartPrec}]
  The construction of $\sem{N}$ from $N = (S,T,F,I)$ starts by building the precubical set $X=\sem{(S, T, F)}$ and then take a subset $X' = \{(m, (t_1,\dotsc, t_n))\in \Nat^S\times \square\mid \forall i=1,\dotsc, n: \forall s\in \prepli{t_i}: m(s)=0\}$.
  Thus we only need to show \eqref{eq:precid-partial}.
  That is if there exist $x_0,x_2 \in X'$ such that $\delta_{C, D}( \delta_{A, B}(x_0)) = x_2$ then $\delta_{A\cup C, B\cup D}(x_0) = x_2$.
  Let $x_0,x_2$ be such cells and let $x_1 \in X'$ such that $\delta_{A, B}(x_0) = x_1$ and $\delta_{C, D}(x_1) = x_2$.
  Assume $x_i=(m_i,U_i)$ and $U_i = (t_{i,1},\dots,t_{i,n_i})$.
  By construction of $X'$ for all $t_{i,j} \in U_i$ and for all $s \in \prepli{t_{i,j}}$, $m_i(s) = 0$.
  As $x_2$ is already in $X'$ it remains to show that $\delta_{A\cup C, B\cup D}(x_0) = x_2$, that is $x_2$ is reachable form $x_0$ by unstarting $A\cup C$ and terminating $B\cup D$.
  By definition of face maps $U_1 = U_0 \setminus (A \cup B)$ and $U_2 = U_1 \setminus  (C \cup D)$.
  Thus $U_2 = U_0 \setminus (A \cup B \cup C \cup D)$.
  Moreover, by construction of $X$, $m_1 = m_0 + \prepla{A} + \prepla{B}$ and $m_2 = m_1 + \pospla{C} + \pospla{D}$.
  Hence $m_2 = m_0 + \prepla{A} + \prepla{B} + \pospla{C} + \pospla{D}$ and $\delta_{A\cup C, B\cup D}((m_0,U_0)) = (m_2,U_2)$.
  \qed
\end{proof}

\end{document}